\documentstyle[pra,aps,graphicx]{revtex}
\begin{document}
\draft
\title{Disorder in Two Dimensional Josephson
 Junctions \footnote{ Phys. Rev. B {\bf 55}, 14499 (1997)}}
\author{ Baruch Horovitz and Anatoly Golub}
 \address{Department of Physics, Ben-Gurion
 University \protect\\ Beer-Sheva, 84105, Israel} 
\maketitle       
\begin{abstract}
An effective free energy of a two dimensional 
(i.e. large area) Josephson Junctions
is derived, allowing for thermal fluctuations, for random magnetic 
fields and for external
currents. We show by using replica
symmetry breaking methods, that the junction has four distinct phases:
disordered, Josephson ordered, a glass phase and a coexisting 
Josephson order with the glass phase.
Near the coexistence to glass transition at $s=1/2$ the 
critical current is $\sim (\mbox{area})^{-s+1/2}$ where $s$ is 
a measure of disorder. 
Our results may account for junction ordering
at temperatures well below the critical temperature of the bulk in 
high $T_c$ trilayer junctions.
\end{abstract} 

 \section{Introduction}

Recent advances in fabrication of Josephson junctions (JJ) have led to
junctions with large area, i.e. the junction length L (in either 
direction in the
junction plane) is much larger than  $\lambda$, the magnetic 
penetration length
in the bulk superconductors. Experimental studies of trilayer 
junctions like
\cite{Rogers} 
$YBa_2Cu_3O_x/PrBa_2Cu_3O_x/YBa_2Cu_3O_x$ (YBCO junction) or like
\cite{Virshup}
 $Bi_2Sr_2CaCu_2O_8/Bi_2Sr_2Ca_7Cu_8O_{20}/Bi_2Sr_2CaCu_2O_8$  (BSSCO
junction) have shown anomalies in the temperature dependence of the 
critical current
$I_c$. In particular in the YBCO junction \cite{Rogers} with area of 
$50 \mu m^2$ a zero resistance state was
achieved only below $~50 K$, although the $YBa_2Cu_3O_x$ layers were 
superconducting already at $T_c\approx 
85 K$. More recent data on similar YBCO junctions 
\cite{Hashimoto,Sato,Strbik} with junction areas of $10^2 - 10^4 \mu 
m^2$ show a measurable $I_c$ only at $20-60 K$ below $T_c$ of the 
superconducting layers. An even larger junction 
\cite{Cucolo} of area $\approx 10^5 \mu m^2$ shows a well defined 
gap structure in the $I-V$ curve while a critical current is not 
observed. In the BSCCO junction \cite{Virshup} a supercurrent through 
the junction could not be observed above $30 K$, although the 
$Bi_2Sr_2CaCu_2O_8$ layer remained
superconducting up to $T_c\approx 80 K$. 

These remarkable observations are significant
both as basic phenomena and for junction applications. In particular, 
these data raise
the question of whether thermal fluctuations or disorder can 
significantly lower the
ordering temperature of two dimensional (2D) junctions.

We note that for both YBCO and BSCCO junctions typically $\lambda 
\approx 0.2 \mu m$ at low temperatures where the junctions order, so 
that the junctions above are 2D in the sense that disorder and 
spatial fluctuations on the scale of $\lambda$ can be important. 
The qualitative effect of these 
fluctuations depends on the Josephson length $\lambda _J$ 
($\lambda _J>\lambda$) which is the width of a Josephson vortex 
(see section II). For $\lambda <L<\lambda _J$ junction parameters 
are renormalized and become L dependent, while more significant 
renormalizations which correspond to 2D phase transitions occur in the 
regime $\lambda_J <L$. From magnetic field dependence \cite{Sato} 
and $L$ dependence \cite{Umezawa} of $I_c$, junctions with 
$\lambda_J <L$
can be realized. The studied junctions are 2D also in the sense the 
thermal fluctuations at temperature T do not lead to uniform large 
phase fluctuations, i.e. $\phi_0 I_c /2c<T$, a condition valid for 
the relevant data (see section V); $\phi_{0}=hc/2e$ is the flux 
quantum.

The energy of a 2D junction, in terms of the Josephson phase 
$\varphi_J(x,y)$
where $(x,y)$ are coordinates in the junction plane, was derived by 
Josephson
\cite{Josephson}. It has the form
\begin{equation}
{\cal F}_0=\int dxdy(\frac{\tau}{16\pi}({\bf\nabla}\phi_J)^2
+\frac{E_J}{\lambda^2} (1-\cos\varphi_J))
                               \label{1}\end{equation}
where $E_J$ is the Josephson coupling energy in area $\lambda^2$.

 Equation~(\ref{1}) was derived \cite{Josephson} on a mean field 
level, i.e.
only its value at minimum is relevant. It was shown, however, (see
Ref.\onlinecite{Horovitz} and Appendix A) that  Eq.~(\ref{1}) 
is valid in a much more
general sense, i.e. it describes thermal fluctuations of 
$\varphi_J(x,y)$ so that
a partition function at temperature $T(<T_c)$
\begin{equation}
Z=\int {\cal D}\phi_J \exp\{-{\cal F}_0[\varphi_J(x,y)]/T\}
\label{2}\end{equation}
is valid.

Equation~(\ref{2}) implies a Berezinskii-Kosterlitz-Thouless 
type phase transition \cite{Kogut} at a temperature $T_J \approx 
\tau$ so that at $T>T_J$
the phase $\phi_J$ is disordered, i.e. the $\cos\phi_J$ correlations 
decay as a power law while at $T<T_J \; \cos \phi_J$ achieves long 
range order. For
the clean system, however, $T_J \approx \tau$ is too close to $T_c$
 for either separating bulk from junction fluctuations 
or for accounting for the experimental data \cite{Horovitz}. 
A consistent description of this
transition, as shown in the present work, can  be achieved by 
allowing for disorder at the
junction, a disorder which reduces $T_J$ considerably.

Equation~(\ref{1}) with disorder is related to a Coulomb gas and 
surface roughening models 
which were studied by replica and
renormalization group (RG) methods \cite{Cardy,Rubinstein}. We find, 
however, that RG
generates a nonlinear coupling between replicas and therefore 
standard replica
symmetric RG methods are not sufficient. In fact, related systems
\cite{Korshunov,Giamarchi} were shown to be unstable towards replica 
symmetry
breaking (RSB).
   
In our system we find a competition between long range Josephson type 
ordering and formation of a glass type RSB phase. The phase
diagram has four phases: a disordered phase, Josephson 
phase (i.e. ordered with finite renormalized
Josephson coupling), a glass phase and  a coexistence phase. The
coexistence phase is unusual in that it has Josephson type long range 
order coexisting
with a glass order parameter. This phase is distinguished from the 
usual ordered phase,
presumably, by long relaxation phenomena typical to glasses 
\cite{Parisi}.

 In the disordered and glass phases fluctuations reduce the critical
current by a power of the junction area,
while in the Josephson and coexistence phases the fluctuation effect 
 saturates when the $(\mbox{area})^{1/2}$ is larger than either the 
Josephson length (in the Josephson phase) or larger than both the
Josephson length and a glass correlation length (in the coexistence 
phase).
These predictions can serve to identify these phases. We show that a
transition between the glass phase and the coexistence phase can
occur well below the critical temperature $T_c$ of the bulk, a result
 which may account for the experimental data on trilayer junctions 
\cite{Rogers,Virshup,Hashimoto,Sato,Strbik}.

In section II we define the model and study the pure case. In section 
III we
study the system with random magnetic fields due to, e.g., quenched 
flux loops in the
bulk and show that RG generates a coupling between different 
replicas. The system with disorder is solved by the
method of one-step RSB \cite{Korshunov,Mezard} in section IV. 
Appendix A derives the free energy of a 2D junction. In particular, 
Appendix A2
allows for space dependent external currents, a situation which, as 
far as we
know, was not studied previously. Appendix B extends the one step 
solution of
section IV to the general hierarchical case, showing that they are 
equivalent.

\section{Thermal effects}

Appendices A.1-A.4 derive the effective free energy of a 2D junction, 
in
presence of an external current $j^{ex}(x,y)$, for the geometry shown 
in Fig.
1. The presence of $j^{ex}(x,y)$ dictates that  the relevant 
thermodynamic
function is a Gibbs free energy, Eq.~(\ref{A10}) which for the
junction becomes (Eqs.~(\ref {A23},\ref{A35}))

\begin{equation}
{\cal G}_J\{\phi_J\}={\cal F}_0\{\phi_J\} - (\phi_0/2\pi c)\int 
dx\,dy\, j^{ex}(x,y)
\varphi_J(x,y) 
                                 \label{3}\end{equation}
where ${\cal F}_0$ is given by 
Eq.~(\ref{1}). The cosine
term is the Josephson tunneling \cite{Josephson} valid for weak 
tunneling
$E_J<<\tau$ and $\tau$ is found in two cases 
(Eqs.~(\ref{A21},\ref{A32})): Case I of long 
superconducting
banks $W_1, W_2>>\lambda$ and case II of short banks, $W_1, 
W_2<<\lambda$,
\begin{equation}
\tau = \left \{ \begin{array}{ll}
\frac{\phi_0^2}{4\pi^2\lambda}  &   \mbox{\hspace{30mm}case I: 
$W_1, W_2>>\lambda$}  \\  \\
\frac{\phi_0^2}{2\pi^2} \frac{W_1W_2}{\lambda_1^2W_2 
+\lambda_2^2W_1}  &
  \mbox{\hspace{30mm}case II: $W_1, W_2<<\lambda$}
                                  \end{array}  \right. 
                       \label{4}\end{equation}
Note that in case II the derivation allows for an asymmetric junction 
with different penetration lengths $\lambda_1, \lambda_2$ and different 
lengths $W_1, W_2$.

It is of interest to note that $j^{ex}$ breaks the symmetry $\varphi_J
\rightarrow \varphi_J+2\pi$, i.e. the external current distinguishes 
between different minima of the cosine term in  Eq.~(\ref{1}). 
For a uniform $j^{ex}$
the Gibbs term reduces to the previously known form \cite{Schon}.

Appendices (A1-A4) present detailed derivation of Eq.~(\ref{3}). This 
derivation
is essential for the following reasons: (i) It shows that the 
fluctuations of
$\varphi_J$ decouple from phase fluctuations in the bulk, (excluding 
flux loops in the bulk which are introduced in section III). 
Thus Eq.~(\ref{3}) is valid below the fluctuation (or
Ginzburg) region around $T_c$. (ii) It shows that Eq.~(\ref{3}) is 
valid for
all configurations of $\varphi_J$ and not just those which solve the 
mean field
equation.

\begin{figure}[htb]
\begin{center}
\includegraphics[scale=0.7]{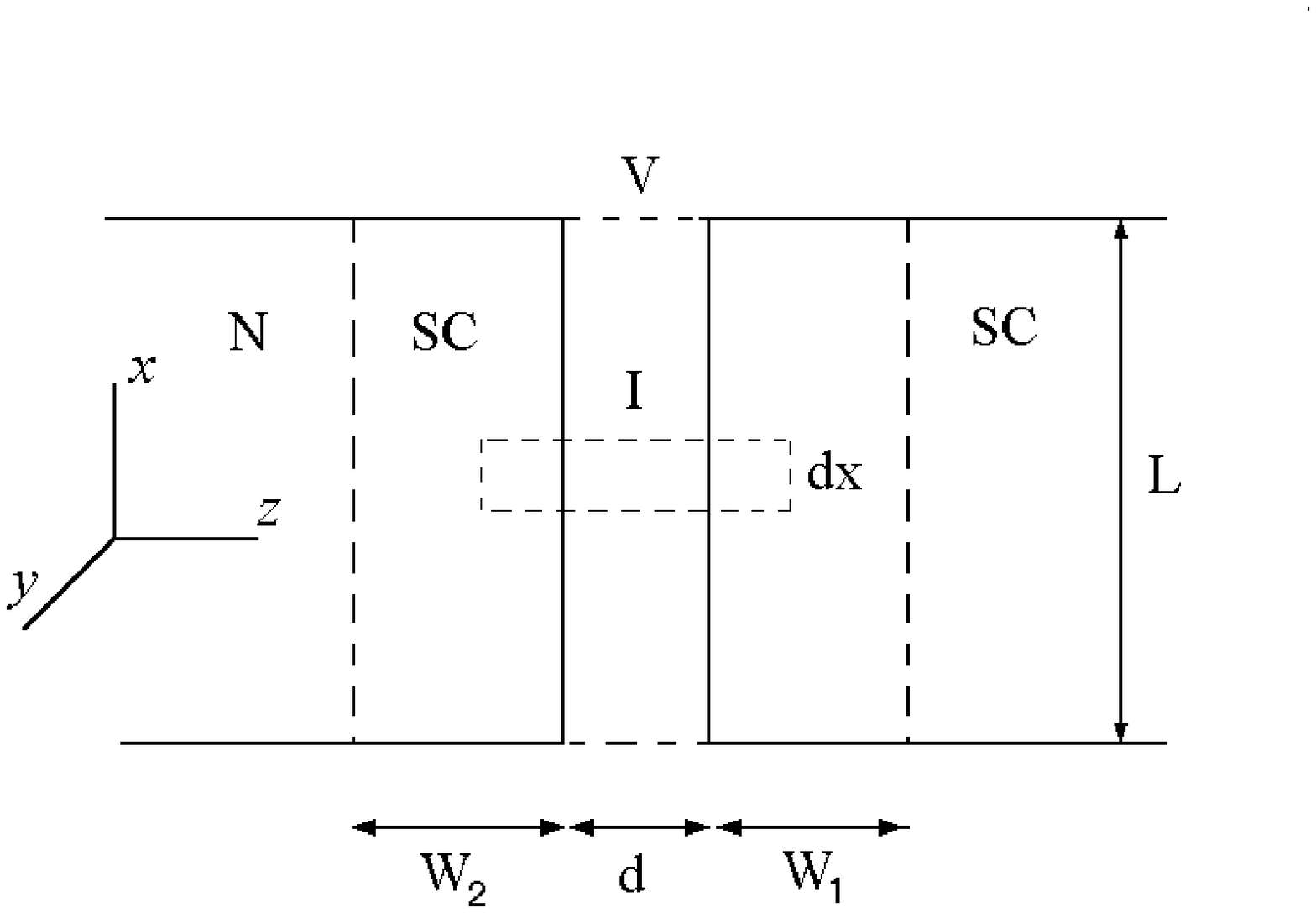}
\end{center}
\caption{Geometry of the 2D Josephson junction. The various components are 
superconductors (S),
insulating barrier (I), normal metal (N) for the external leads and 
vacuum (V). The
dashed rectangle serves to derive boundary conditions in Appendix A 1.}
\end{figure}

It is instructive to consider the mean field equation
$\delta G_J/\delta\varphi_J=0$, i.e.
\begin {equation}
\frac{E_J}{\lambda^2}\sin\varphi_J = \frac{\tau}{8\pi}{\bf\nabla}^2 
\varphi_J +
\frac{\phi_0}{2\pi c}j^{ex}
                            \label{5}\end{equation}

This equation can also be derived by equating the current
$j_z = (-c/4\pi\lambda^2) A_z'$ at $z=d/2$ (given, e.g. in case I by
Eqs.~(\ref{A17}, \ref{A20})] with the Josephson tunneling current
$j_J=(2\pi c/\phi_0)(E_J/\lambda^2)\sin\varphi_J$. Eq.~(\ref{5}), 
however, is
not on a level of conservation law or a boundary condition since
configurations which do not satisfy Eq.~(\ref{5}) are allowed in the 
partition
sum. More generally,  Eq.~(\ref{5}) is satisfied only after thermal 
average
$\langle\delta G_J/\delta \varphi_J\rangle =0$. An equivalent way of 
studying
thermal averages is to add to Eq.~(\ref{5}) time dependent 
dissipative and
random force terms. The time average, which {\em includes} 
configurations which do
not satisfy Eq.~(\ref{5}), is by the ergodic hypothesis equivalent to 
the
partition sum, i.e. a functional integral over $\varphi_J$ with the 
weight
$\exp[-G_J/T]$.

Eq.~(\ref{1}) is the well known 2D sine-Gordon system \cite{Kogut} 
which for
$j^{ex}=0$ exhibits a phase transition. Since renormalization group 
(RG)
proceeds by integrating out rapid variations in $\varphi_J$, 
$j^{ex}\neq 0$ is
not effective if it is slowly varying (e.g. as in case II).

RG integrates fluctuations of $\varphi_J$ with wavelengths between 
$\xi$ and
$\xi+\,d\xi$, the initial scale being $\lambda$. The parameters 
$t=T/\tau$ and
$u=E_J/T$ are renormalized, to second order in $u$, via \cite{Kogut}
\begin{eqnarray}
du/u&=&2(1-t)\,d\xi/\xi  \nonumber\\
dt&=&2\gamma^2 u^2 t^3\,d\xi/\xi
\label{6}\end{eqnarray}                                  
where $\gamma$ is of order 1 (depending on the cutoff smoothing 
procedure). 
Eq.~(\ref{6}) defines a phase transition at $1/t=1-\gamma u$.
Note, however, that $\tau$ itself is temperature dependent since
$\lambda(T)=\lambda'(1-T/T_c^0)^{-1/2}$, where $T_c^0$ is the mean 
field
temperature of the bulk. Thus the solution of $\tau(T)/T=1-\gamma 
E_J/T$
defines a transition temperature $T_J$ which is below $T_c^0$. 
However, $T_J$
is too close to $T_c^0$ and is in fact within the Ginzburg 
fluctuation region
around $T_c^0$. To see this, consider a complex order parameter
$\psi=|\psi|\exp(i\varphi)$ with a free energy of the form

\[{\cal F}=\int d^3r[a|\psi|^2+b|\psi|^4+a\xi^2|{\bf \nabla}\psi|^2] 
\]

The Ginzburg criterion equates fluctuations with $b=0$, i.e. $\langle
\delta\psi^2\rangle\approx T/a\xi^3$ with $|\psi|^2(=|a|/2b)$ in the 
ordered
phase. Since $|{\bf \nabla}\psi|^2\approx|\psi|^2({\bf 
\nabla}\varphi)^2$
Eq.~(\ref{A13}) identifies 
$a\xi^2|\psi|^2=(\phi_0/2\pi\lambda)^2/8\pi$, 
so that the Ginzburg temperature is
\begin{equation}
T^{Ginz}=a\xi^3|\psi|^2=\xi(\phi_0/2\pi\lambda)^2/8\pi.
               \label{7}\end{equation}

Since $\xi<\lambda, W$ in both cases I and II, $T^{Ginz}<T_J$. The 
neglect of
flux loop fluctuations, as assumed in appendices A3, A4 is therefore 
not justified at $T_J$. Thus the relevant range of temperatures for 
the free energy Eqs.~(\ref{1},\ref{3}) is $T\ll T^{Ginz}<\tau$, i.e. 
$t\ll 1$.

   The RG Eqs.~(\ref{6}) can, however, be used in the range $T<T_J$ to
study fluctuation effects in the ordered region. Excluding a narrow 
interval
near $T_J$ where $|\tau/T-1|<\gamma E_J/T<<1$ renormalization of $t$ 
can be
neglected and integration of Eq.~(\ref{6}) yields a renormalized 
Josephson
coupling $E_J^R=E_J(\xi/\lambda)^{2(1-t)}$. Scaling stops at the
Josephson length $\lambda_J$ at which the coupling becomes strong,
$E_J^R\approx\tau/8\pi$ (the $8\pi$ is chosen so that 
$\lambda_J=\lambda_J^0$ at $T=0$, where $\lambda_J^0$ is the conventional 
Josephson length). Thus
$\lambda_J=\lambda(\tau/8\pi E_J)^{1/[2(1-t)]}$; the $T=0$ value is
$\lambda_J^0=\lambda(\tau/8\pi E_J)^{1/2}$. The scaling process is 
equivalent
to replacing $(E_J/\lambda^2)\langle\cos \varphi_J\rangle$ by
$\tau/8\pi\lambda_J^2$ so that $\langle\cos
\varphi_J\rangle=(\lambda_J^0/\lambda_J)^2$ is the reduction factor
due to fluctuations.

The free energy Eqs.~(\ref{1},\ref{3}) with renormalized parameters 
yields a critical current by a mean field equation (see comment 
below Eq.~(\ref{9})). The renormalized junction is
either an effective point junction ($L<\lambda_{J}$) with the current 
flowing through the whole junction area, or a strongly coupled 
($E_{J} \approx \tau /8\pi$) 2D junction where the current 
flows near the edges of the
junction with an effective area $L\lambda_J$. The mean field critical 
currents \cite{Kulik}  are
\begin{eqnarray}
I_{c1}^0&=& (2\pi c/\phi_0)E_J(L/\lambda)^2  \mbox{\hspace{20mm}}  
L<\lambda_J      \nonumber\\ 
I_{c2}^0&=& c\tau L/2\phi_0\lambda_J^0  \mbox{\hspace{32mm}}   
L>\lambda_J
     \label{7a}\end{eqnarray}

 The effect of fluctuations is to
reduce $E_J$ so that the critical current is
\begin{equation}
I_c=I_{c1}^0\,(L/\lambda)^{-2t} \mbox{\hspace{33mm}$L<\lambda_J$} .
                                    \label{8}\end{equation}

 In the second case, $L>\lambda_J$, the fluctuations reduce the 
current density by $\langle\cos\varphi_J\rangle$ but enhance 
the effective area by
$\lambda_J/\lambda_J^0=\langle\cos\varphi_J\rangle^{-1/2}$. The 
critical current is then
\begin{equation}
I_c=I_{c2}^0(4\pi E_J/\tau)^{t/[2(1-t)]} 
\mbox{\hspace{27mm}$L>\lambda_J$} .
                             \label{9}\end{equation}
Thus even if $t<<1$ in Eqs.~(\ref{8},\ref{9}) a sufficiently small 
$E_J$ can lead to
an observable reduction of $I_c$.

Note that thermal fluctuations act to renormalize $E_J$ which then 
determines a critical current by the mean field equation. This 
neglects thermal fluctuations in which $\varphi_J$ fluctuates 
uniformly over the whole junction. These fluctuations can be 
neglected when the coefficient of the cosine term in Eq.~(\ref{1}) 
(including the area integration) is larger then temperature, i.e. in 
terms of $I_c$, $\phi_0 I_c /2c >T$. This condition is consistent 
with experimental data (see section V).

\section{Disorder and RG}

There are various types of disorder in a large area junction. 
An obvious type are spatial variations in the Josephson coupling 
$E_J$. A random distribution of $E_J$ with zero mean is equivalent 
to known systems
\cite{Korshunov,Giamarchi} and produces only a glass phase. The more 
general situation is
to allow a finite mean of $E_J$, and allow for another type of 
disorder, i.e. random
coupling to gradient terms. Since the magnetization of the junction 
is proportional to \cite{Josephson}
${\bf \nabla }\varphi_J$ we propose that the most 
interesting type of
disorder are random magnetic fields. Such fields can arise from 
magnetic impurities, or
more prominently from random flux loops in the bulk.

A flux loop in the bulk with radius $r_0$ has a magnetic field of 
order $\phi_0/2\pi\lambda^2$ in the vicinity of the loop. A 
straightforward solution
of London's equation shows that the field far from the loop depends 
on the ratio
$r_0/\lambda$. For large loops, $r_0>\lambda$, the field at distance 
$r>>r_0$ decays
exponentially while for small loops $r_0<<\lambda$, it decays slowly 
as $1/r^2$ 
($\lambda>r>>z, r_0$, where $r$ is in the loop plane and $z$ is 
perpendicular to it) or
as $1/z^3$ ($\lambda>z>>r, r_0$). Thus, the local magnetic field has 
contributions from all
flux loops of sizes $r_0<\lambda$. If $P(r_0)$ is the probability of 
having a flux loop of
size $r_0$ then the local average magnetic field is of order 
\begin{equation}
H_s^2 \approx [(\phi_0/2\pi\lambda^2)\int^{\lambda} P(r_0)\,dr_0]^2 
\equiv 4s\phi_0^2/\pi\lambda^4
                 \label{10}\end{equation}

The last equality defines a measure of disorder $s$ which
increases with the $r_0$ integration, say as $s\sim\lambda^\alpha$ 
with $\alpha>0$.
The distribution of $H_s$ is therefore of the form $\exp
[-\pi H_s^2\lambda^4/4s\phi_0^2]$.

Consider a dimensionless random field ${\bf q}(x,y)=\lambda 
\sqrt{8\pi}{\bf
H}_s(x,y)/4\phi_0$ so that its distribution is
\begin{equation}
\exp[-\lambda^2\sum_{x,y}{\bf q}^2(x,y)/2s]=\exp[-\int{\bf
q}^2(x,y)\,dxdy/2s]
                              \label{11}\end{equation}

The coupling of magnetic fields to the Josephson phase is from 
Eqs.~(\ref{A20},\ref{A39}))
and for $\tau$ of case I (Eq.~(\ref{4}))
\begin{equation}
{\cal F}_s=-(\tau/\sqrt{8\pi})\int\,dx\,dy\,(\hat{\bf z}\times {\bf
\nabla}\varphi_J(x,y))\cdot{\bf q}(x,y)
                                         \label{12}\end{equation}

The fields in Eq.~(\ref{10}) are in fact relevant only to case I. In 
case II image
flux loops across the superconducting-normal (SN) surface reduce the 
contribution of loops
with $r_0<W$. Thus  Eq.~(\ref{10}) is valid with the $r_0$ 
integration limited by $W$.
Since now $\tau=\phi_0^2W/4\pi^2\lambda^2$ (Eq.~(\ref{4}) for 
symmetric junction) we define
${\bf q}(x,y)=\sqrt{8\pi}\lambda^2{\bf H}_s(x,y)/4\phi_0W$ so that 
the coupling 
Eq.~(\ref{12}) has the same form. The distribution of ${\bf q}(x,y)$ 
has the same
form as in Eq.~(\ref{11}) except that now $s\sim\lambda^2$. Since 
$\lambda$ is $T$
dependent, $s$ is also $T$ dependent, a feature which is relevant to 
the experimental data (see section V).

We proceed to solve the random magnetic field problem by the replica 
method \cite{Parisi}. We raise the partition sum to a power $n$, 
leading to replicated
Josephson phases $\varphi_\alpha$, $\alpha=1,...,n$. The factor ${\bf 
q}(x,y)$ in
Eq.~(\ref{12}) is then integrated with the weight Eq.~(\ref{11}), 
leading to

\begin{equation}
Z^{(n)}\sim\exp[(s\tau^2/16\pi T^2)(\sum_{\alpha}{\bf
\nabla}\varphi_{\alpha})^2]  .
                                      \label{13}\end{equation}
 
In this section we attempt to solve the system by RG methods 
\cite{Cardy,Rubinstein}. We
find, however, that RG generates nonlinear couplings between replicas 
which
eventually lead to replica symmetry breaking (section IV). Thus the 
direct
application of RG is not sufficient.

Consider first the Gaussian part 
\begin{equation}
{\cal F}_0^{(n)}=\frac{1}{2}\int dx\,dy\sum_{\alpha, \beta}M_{\alpha, 
\beta}{\bf \nabla}\varphi_{\alpha}{\bf \nabla}\varphi_{\beta}                                                     
\label{14}\end{equation}
with
\begin{equation}
M_{\alpha , \beta}=\frac{1}{8\pi t}\delta _{\alpha , 
\beta}-\frac{s}{8\pi t^2} .
                                          \label{15}\end{equation}
(From here on T is absorbed in the definition of free energies, i.e.
 ${\cal F}\rightarrow {\cal F}/T)$.

We use Eq.~(\ref{14}) to test for relevance of terms of the form
$v^{(\ell)}\cos(\sum_{i=1}^{\ell}\eta_i\varphi_{\alpha_i})$. These 
terms are generated from
powers of the $\sum_{\alpha}\cos\varphi_{\alpha}$ interaction in 
presence of the disorder
$s$. First order RG is obtained by integrating a high momentum field 
$\zeta_{\alpha}$ with
momentum in the range $\xi^{-1}+d(\xi^{-1})<q<\xi^{-1}$. The Green's 
function, averaged
over these high momentum terms in Eq.~(\ref{14}), is
\begin{eqnarray}
G_{\alpha, \beta}({\bf r})=\langle\zeta_{\alpha}({\bf
r})\zeta_{\beta}(0)\rangle&=&(M^{-1})_{\alpha, \beta}\int 
d^2q\,\exp(-i{\bf
q}\cdot{\bf r})/(2\pi q)^2 \nonumber\\
   &=& (M^{-1})_{\alpha, \beta}J_0(r/\xi)d\xi/2\pi\xi.
                                     \label{16}\end{eqnarray}
Defining $\varphi_{\alpha}=\chi_{\alpha}+\zeta_{\alpha}$, RG to first 
order is
obtained by integrating $\zeta_{\alpha}$,
\begin{eqnarray}
&\sum_{\bf  
r}\langle\cos(\sum_{i=1}^{\ell}\eta_i\varphi_{\alpha_i})\rangle =
\sum_{\bf r}\cos(\sum_{i=1}^{\ell}\eta_i\chi_{\alpha_i})
\exp[-\frac{1}{2}\langle
(\sum_{i=1}^{\ell}\eta_i\zeta_{\alpha_i})^2\rangle]  \nonumber\\ 
&=\sum_{\bf
r}'\cos(\sum_{i=1}^{\ell}\eta_i\chi_{\alpha_i})[1+2\frac{d\xi}{\xi}
-\frac{m}{2}G_1(0)-\frac{1}{2}\sum_{i\neq j}\eta_i\eta_jG_2(0)]
                                      \label{17}\end{eqnarray}
where $\sum '$ denotes summation on a unit cell larger by 
$1+2d\xi/\xi$ and
\begin{eqnarray}
G_1(0)&=&G_{\alpha,\alpha}(0)=\left(8\pi
t+\frac{8\pi s}{1-ns/t}\right)\frac{d\xi}{2\pi\xi}
                       \nonumber\\
G_2(0)&=&G_{\alpha \neq \beta}(0)=\frac{8\pi 
s}{1-ns/t}\frac{d\xi}{2\pi\xi}
                                  \label{18}\end{eqnarray}

The most relevant operators in Eq.~(\ref{17}) are when $\sum_{i\neq 
j}\eta_i\eta_j$ is minimal, i.e. $\sum_i \eta_i=0$ for even $\ell$ or 
$\sum_i \eta_i=\pm 1$ for odd $\ell$. Thus,
\begin{eqnarray}
dv^{(\ell)}&=& 2v^{(\ell)}(1-\ell t)d\ln \xi    
\mbox{\hspace{34mm}$\ell$
even}      \nonumber\\
 dv^{(\ell)}&=&2v^{(\ell)}(1-\ell t-s)d\ln \xi  
\mbox{\hspace{28mm}$\ell$ odd}
\label{19}\end{eqnarray}                     
Thus, as temperature is lowered, successive $v^{(\ell)}$ terms become 
relevant at
$t<1/\ell$ ($\ell$ even) and at $t<(1-s)/\ell$ ($\ell$ odd).

We consider in more detail the $v=v^{(2)}$ term, the lowest order 
term which mixes
different replica indices. The free energy of this model has the form

\begin{equation}
{\cal F}^{(n)}=\int dx\,dy\{\frac{1}{2}\sum_{\alpha, \beta}M_{\alpha, 
\beta}{\bf \nabla}\varphi_{\alpha}{\bf \nabla}\varphi_{\beta}
-\frac{u}{\lambda^2}\sum_{\alpha}\cos\varphi_\alpha -
\frac{v}{\lambda^2}\sum_{\alpha, \neq 
\beta}\cos(\varphi_{\alpha}-\varphi_{\beta})\}
                                \label{20}\end{equation}

Note that the $v$ term is also generated by disorder in the Josephson 
coupling, corresponding to a distribution with a mean value $\sim u$. 
If $u=0$ Eq.~(\ref{20})
reduces to the well studied case \cite{Korshunov,Giamarchi} with a 
glass phase a low
temperatures. We consider here the more general case of $u\neq 0$, 
which indeed leads to
a much more interesting phase diagram.

 The initial values for RG flow are $u=E_J/T, v=0$.
Standard RG methods \cite{Kogut} to second order in $u, v$ lead to 
the following set of
differential equations: 
\begin{eqnarray} 
du&=&[2u(1-t-s)-2\gamma'yvt]d\ln\xi  
\nonumber\\ dv&=&[2v(1-2t)+(1/2)\gamma's u^2-2\gamma 'tv^2]d\ln\xi   
\nonumber\\
dt&=&-2\gamma^2(t+s)t^2u^2d\ln\xi           \nonumber\\ 
d(s/t^2)&=&16\gamma
^2tv^2d\ln\xi
                                          \label{21}\end{eqnarray}
where $\gamma ,\gamma '$ are numbers of order 1 (depending on cutoff 
smoothing procedure).
  
Note that any $u\neq 0$ generates an increase in $v$, so that $v=0$ 
cannot be a fixed
point. In contrast, $v\neq 0$ allows for a $u*=0$ fixed point 
(ignoring for a moment
the flow of $s$), with $u*=0,\; v*=(1-2t)/\gamma 't$. This fixed 
point is stable in the $(u, v)$ plane if $t<1/2,\, s$; however, $s$ 
increases without bound. This indicates that the $v$ term is essential 
for the behavior of the system.

We do not explore Eq.~(\ref{21}) in detail since it assumes replica 
symmetry, i.e.
the coefficient $v$ is common to all $\alpha, \beta$. In the next 
section we show that the system favors to break this symmetry, 
leading to a new type of ordering.

\section{replica Symmetry Breaking}

The possibility of replica symmetry breaking (RSB) has been studied 
extensively in
the context of spin glasses \cite{Parisi} and applied also to other 
systems. In
particular, the free energy Eq.~(\ref{20}) with $u=0$ was studied in 
the context of
flux line lattices and of an XY model in a random field 
\cite{Korshunov,Giamarchi}.
In this section we use the method of one-step replica symmetry 
breaking
\cite{Korshunov,Mezard} for the Hamiltonian Eq.~(\ref{20}); in 
appendix B we present
the full hierarchical solution, which for our system turns out to be 
equivalent to
the one-step solution.

Consider the self consistent harmonic approximation \cite{Korshunov} 
in which one finds a Harmonic trial hamiltonian

\begin{equation}
{\cal H}_0=\frac{1}{2}\sum_q\sum_{\alpha ,\beta}\,G^{-1}_{\alpha,
\beta}(q)\varphi_{\alpha}\varphi_{\beta}^*(q)                                               
\label{22}\end{equation}
such that the free energy
\begin{equation}
{\cal F}_{var}={\cal F}_0+\langle{\cal H}-{\cal H}_0\rangle_0
                                          \label{23}\end{equation}
is minimized. ${\cal H}={\cal F}^{(n)}/T$ is the interacting 
Hamiltonian,
 Eq.~(\ref{20}), ${\cal F}_0$ is the free energy corresponding to 
${\cal H}_0$ and
$\langle...\rangle_0$ is a thermal average with the weight 
$\exp(-{\cal H}_0)$. The
interacting terms lead to 

\begin{eqnarray}
\int d^2r\,\langle\cos\varphi_{\alpha}({\bf 
r})\rangle_0&=&\exp(-A_{\alpha}/2)
\nonumber\\ A_{\alpha}&=&\sum_q\langle|\varphi_{\alpha}(q)|^2\rangle=
\sum_qG_{\alpha,\alpha}       \nonumber\\
\int d^2r\,\langle\cos (\varphi_{\alpha}-\varphi_{\beta})\rangle_0&=&
\exp(-B_{\alpha,\beta}/2) \nonumber\\ 
B_{\alpha,\beta}=\sum_q\langle|\varphi_{\alpha}(q)-
\varphi_{\beta}(q)|^2\rangle 
 &=&\sum_q[G_{\alpha,\alpha}+G_{\beta,\beta}-G_{\alpha 
,\beta}-G_{\beta ,\alpha}]  
                               \label{24}    \end{eqnarray}

Therefore

\begin{eqnarray}
{\cal F}_{var}=&-\frac{1}{2}\sum Tr[\ln
\hat{G}(q)+(\hat{G}^{-1}(q)+\hat{M}q^2)\hat{G}(q)] \nonumber\\
-&\frac{u}{\lambda^2}\sum_{\alpha}\exp(-\frac{1}{2}A_{\alpha}) 
-\frac{v}{\lambda^2}\sum_{\alpha\neq\beta}\exp(-\frac{
1}{2}B_{\alpha ,\beta})
                                      \label{25}\end{eqnarray}
where the $Tr\,\ln\hat{G}(q)$ term corresponds to ${\cal F}_0$ (up to 
an additive
constant) and the $\hat{}$ sign denotes a matrix in replica space.

We define now $u_0=8\pi tu/\lambda^2$, $v_0=16\pi tv/\lambda^2$ and 
using Eq.~(\ref{15}) the minimum condition $\delta{\cal F}_{var}/
\delta G_{\alpha ,\beta}=0$ becomes
\begin{equation}
\hat{G}(q)=8\pi t[(q^2+u_0\exp(-\case{1}{2}A_{\alpha}))\hat{I}
-\frac{s}{t}q^2\hat{L}-v_0\hat{\sigma}]^{-1}                                               
\label{26}\end{equation}
where $\hat{I}$ is the unit matrix, $\hat{L}$ is a matrix with all 
entries $=1$.
i.e. $L_{\alpha ,\beta}=1$, and $\hat{\sigma}$ is given by
\begin{equation}
\sigma_{\alpha ,\beta}=\exp(-\case{1}{2}B_{\alpha 
,\beta})-\delta_{\alpha
,\beta}\sum_{\gamma}\exp(-\case{1}{2}B_{\alpha ,\gamma}).                                                    
\label{27}\end{equation}
Note that the sum on each row vanishes, $\sum_{\beta}\sigma_{\alpha 
,\beta}=0$.

Consider first briefly the replica symmetric solution. A single 
parameter $\sigma_0$
defines $\hat{\sigma}$ so that the constraint 
$\sum_{\beta}\sigma_{\alpha ,\beta}=0$ yields
\begin{equation}
\hat{\sigma}=\sigma_0\hat{L}-n\sigma_0\hat{I}                                                  
\label{28}\end{equation}
Using $\hat{L}^2=n\hat{L}$ it is straightforward to find the inverse 
in Eq.~(\ref{26}). In terms of an order parameter 
$z=u_0\exp(-A_{\alpha}/2)$, Eq.~(\ref{27}) with $n\rightarrow 0$ yields 
$\sigma_0=(z/\Delta_c)^{2t}$ where
$\Delta_c (\approx 1/\lambda^2)$ is a cutoff in the $q^2$ integration 
so that
$z<<\Delta_c$ is assumed. The definition of $z$ yields
\[z=u_0\left(\frac{z}{\Delta_c}\right)^{t+s}\exp(s-tv_0\sigma_0/z)\]
                                         
For $t v_0\sigma_0/z<<1$ a consistent $z<<\Delta_c$ solution is 
possible at
$t<1-s$. (Indeed $tv_0\sigma_0/z<<1$ since $\sigma_0<<1$, except at
$z \rightarrow 0$, i.e. at $t \rightarrow 1-s$.) Hence, (neglecting an
$\exp(s)$ factor)
\begin{equation}
z/\Delta_c \approx (u_0/\Delta_c)^{1/(1-t-s)}
                            \label{29}\end{equation}
The replica symmetric solution thus reproduces the 1st order RG 
solution 
(Eq.~(\ref{19}) with $\ell=1$). The order parameter $z$ corresponds 
to $1/\lambda_{J}^2$ of
 Eq.~(\ref{19}) where the Josephson length $\lambda_{J}$ is the scale 
 at which strong coupling is achieved,
$v^{(1)}(\lambda_{J})\approx 1$, and RG stops.

Consider now  a one-step RSB solution of the form 
\cite{Korshunov,Mezard}
\begin{equation}
\hat{\sigma}=\sigma_0\hat{L}+(\sigma_1-\sigma_0)\hat{C}-[\sigma_0n+
m(\sigma_1-\sigma_0)]\hat{I}
                                          \label{30}\end{equation}
where $\hat{C}$ is a matrix with entries of $1$ in $m\times m$ 
matrices which touch
along the diagonal and $0$ otherwise; m is treated as a variational 
parameter. The
coefficient of $\hat{I}$ is fixed by the constraint 
$\sum_{\beta}\sigma_{\alpha ,\beta}=0$.

Eq.~(\ref{30}) corresponds to two order parameters,
\begin{eqnarray}
z&=&u_0 \exp(-A_{\alpha}/2)          \nonumber\\
\Delta&=&v_0[\sigma_0n+m(\sigma_1-\sigma_0)]
                                           \label{31}\end{eqnarray}
The inverse matrix in Eq.~(\ref{26}) is obtained by using 
$\hat{L}^2=n\hat{L}$,
 $\hat{C}\hat{L}=m\hat{L}$ and $\hat{C}^2=m\hat{C}$. It has the form
\begin{equation}
\hat{G}=[a(q)\hat{I}+b(q)\hat{L}+c(q)\hat{C}]^{-1}=\alpha(q)\hat{I}+
\beta(q)\hat{L}+\gamma(q)\hat{C}
                                   \label{32}\end{equation}
and is solved by
\begin{eqnarray}
\alpha(q)&=&1/a(q)                         \nonumber\\
\beta(q)&=&-b(q)[a(q)+mc(q)]^{-1}[a(q)+nb(q)+mc(q)]^{-1}   \nonumber\\
\gamma(q)&=&\{-a^{-1}(q)+[a(q)+mc(q)]^{-1}\}/m
                                           \label{33}\end{eqnarray}

Identifying $a(q), b(q), c(q)$ from Eqs.~(\ref{26},\ref{30}) we 
obtain (after
$n\rightarrow 0$) 
\begin{eqnarray}
\alpha&\equiv &\sum_q\alpha(q)=2t\ln [\Delta_c/(z+\Delta)]          
\nonumber\\
\beta&\equiv 
&\sum_q\beta(q)=2s\ln(\Delta_c/z)+(2/z)tv_0\sigma_0-2s  
\nonumber\\ \gamma&\equiv &\sum_q\gamma(q)=-(2t/m)\ln[z/(z+\Delta)]                                                      
\label{34}\end{eqnarray}
The definitions of $\hat{\sigma}$ and $z$ identifies
the parameters 
\begin{eqnarray}
\sigma_1&=&\exp(-\alpha)                   \nonumber\\
\sigma_0&=&\exp(-\alpha-\gamma)             \nonumber\\
z&=&u_0\exp[-(\alpha+\beta+\gamma)/2]
                                      \label{35}\end{eqnarray}

These equations determine the order parameters $z$, $\Delta$ in terms 
of $m$ and
the parameters of the hamiltonian. The value of $m$ must be 
determined by
minimizing the free energy ${\cal F}_{var}$. (However, in the 
hierarchical scheme with $\Delta (m)$ as {\em function} of $m$,
the variation with respect
to $G_{\alpha ,\beta}$ is sufficient to determine the position of a 
step in $\Delta (m)$, see Appendix B).

Consider first the Gaussian terms ${\cal F}_3$, i.e. the trace term in
Eq.~(\ref{25}). Since this term contains the uninteresting vacuum 
energy
($z=\Delta=0$) it is useful to find the differential $d{\cal F}_3$ 
and then
integrate. Using Eq.~(\ref{32}) for $d\hat{G}(q)$ we have
\begin{equation}
d{\cal 
F}_3=-\frac{1}{2}\sum_qTr\{[\hat{G}^{-1}(q)-\hat{M}q^2][\hat{I}\,d\alpha(q)
+\hat{L}\,d\beta(q)+\hat{C}\,d\gamma(q)]\}                                               
\label{36}\end{equation}
Performing the trace and expressing $d\alpha, d\gamma$ in terms of 
$dz, d\Delta$
(from Eq.~(\ref{34})) we obtain for the free energy per replica, 
$f={\cal F}^{(n)}/n$,
\begin{equation}
df_3=(1-\frac{1}{m})\,d(z+\Delta)+(\frac{z}{m}-v_0\sigma_0)\frac{dz}{z}
-\frac{z}{2t}\,d\beta
                           \label{37}\end{equation}
Integrating $\partial f_3(z,\Delta')/\partial\Delta'$ from 0 to 
$\Delta$, and then
$\partial f_3(z',0)/\partial z'$ from 0 to $z$ adds up to
\begin{equation}
8\pi[f_3(z,\Delta)-f_3(0,0)]=(1-1/m)\Delta-v_0\exp[-\alpha(z,\Delta)-
\gamma(z,\Delta)]+(1+s/t)z  \,.
                                            \label{38}\end{equation}
 The $u$ and $v$ terms in Eq.~(\ref{25}) lead, by using 
Eq.~(\ref{24}), to
$\sim\exp [-(\alpha +\beta +\gamma)]$ and to
$\sim\sum_{\alpha}\sigma_{\alpha,\alpha}=[\sigma_1-(\sigma_1-\sigma_0)m]$,

respectively. Finally, we have
\begin{eqnarray}
&8\pi f(z,\Delta)=8\pi f(0,0)+(1-\frac{1}{m})\Delta+(1+\frac{s}{t})z-
v_0(1-\frac{m}{2t})\mbox{e}^{-\alpha-\gamma} \nonumber\\
&+\frac{v_0}{2t} 
(1-m)\mbox{e}^{-\alpha}-\frac{u_0}{t}\mbox{e}^{-\frac{1}
{2}[\alpha+\beta+\gamma]}
                                          \label{39}\end{eqnarray}
where $\alpha$, $\beta$, $\gamma$ are functions of $z$ and $\Delta$ 
from
Eq~(\ref{34}). Since Eqs.~(\ref{35}) are already minimum conditions, 
it must be
checked that $\partial f/\partial z=\partial f/\partial \Delta=0$ 
reproduces these
equations so that $m$ in Eq.~(\ref{39}) can be taken as an 
independent variational
parameter. The latter statement is indeed correct and $\partial 
f/\partial m=0$
leads to the relation
\begin{equation}
m=\frac{2t\Delta+2tz\ln[z/(z+\Delta)]}{\Delta 
+2tv_0\sigma_0\ln[z/(z+\Delta)]}
                                       \label{40}\end{equation}
Rewriting Eq.~(\ref{35}) with Eq.~(\ref{34}), we have the following 
relations:

\begin{equation}
z=u_0e^{s}\left(\frac{z}{\Delta_c}\right)^{s+t/m}\left(\frac{z+\Delta}
{\Delta_c}\right)^{t(1-1/m)}e^{-tv_0\sigma_0/z}
                              \label{41}\end{equation}
\begin{equation}
\Delta=v_0m\left(\frac{z+\Delta}{\Delta_c}\right)^{2t}\left[1-\left(\frac{z}
{z+\Delta}\right)^{2t/m}\right]
                                          \label{42}\end{equation}
\begin{equation}
\sigma_0=\left(\frac{z+\Delta}{\Delta_c}\right)^{2t}\left(\frac{z}{z+\Delta}
\right)^{2t/m}
                                            \label{43}\end{equation}

\begin{figure}[htb]
\begin{center}
\includegraphics[scale=0.7]{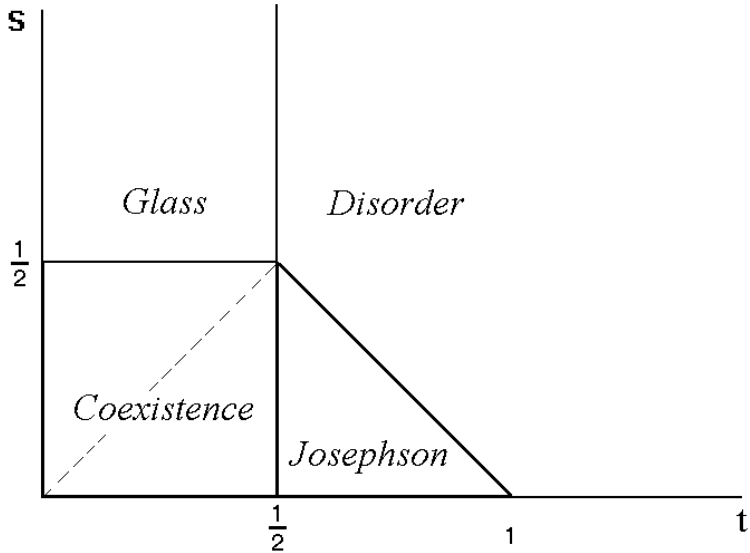}
\end{center}
\caption{ Phase diagram of a 2D junction in terms of s, the spread in random 
magnetic fields and
$t$, which is proportional to temperature. The various phases, in terms 
of the Josephson
order $z$ and the glass order $\Delta$ are: (i) Disordered phase with 
$z=\Delta=0$, (ii) Josephson phase with $z\neq 0,\;\Delta=0$, 
(iii) coexistence with both $z\neq 0,\;\Delta\neq 0$, and (iv) 
glass phase with $z=0,\;\Delta\neq 0$. The dashed line within the 
coexistence phase is where $\Delta$ changes sign.}
\end{figure}

The solutions for $z$ and $\Delta$ of Eqs.~(\ref{40}-\ref{43}) 
determine the phase
diagram. Consider first the Josephson ordered phase $z\neq0, 
\Delta=0$. Expecting
$\sigma_0<<1$ an expansion of Eq.~(\ref{40}) in powers of $\Delta/z$ 
yields
$m\approx t\Delta/z$ so that $\sigma_0\approx e^2(z/\Delta_c)^{2t}$ 
is indeed small.
The solution for $z$ when $\Delta\rightarrow 0$ is equivalent to the 
replica
symmetric case, Eq.~(\ref{29}) and is possible for $t<1-\sigma$. 

Consider next an RSB solution $z=0, \Delta\neq 0$. Eq.~(\ref{40}) 
yields $m=2t$ and Eq.~(\ref{42}) leads to
\begin{equation}
\Delta/\Delta_c=(2tv_0/\Delta_c)^{1/(1-2t)}
                                           \label{44}\end{equation}
Thus a glass type phase is possible for $t<1/2$. (Curiously, a similar
result is obtained for the $v$ term in 1st order RG, $\ell=2$ in 
Eq.~(\ref{19}),
however, $G_{\alpha,\alpha}\sim 1/q^4$ at $q\rightarrow 0$, while here
$G_{\alpha,\alpha}\sim 1/q^2$).

Finally consider a coexistence phase, where both $z$, $\Delta\neq 0$. 
It is remarkable
that $m=2t$ is an exact solution even in this case, as can be checked 
by substitution
in Eqs.~(\ref{40},\ref{42},\ref{43}). The resulting solutions are
\begin{eqnarray}
\frac{z+\Delta}{\Delta_c}&=&\left(2t\frac{v_0}{\Delta_c}\right)^{1/(1-2t)} 
\nonumber\\
\frac{z}{\Delta_c}&=&\mbox{e}^{-1}\left(\frac{u_0^2}{2tv_0\Delta_c}\right)
^{1/(1-2s)}\,.
                                  \label{45}\end{eqnarray}
This coexistence phase is therefore possible at $t<1/2$ and $s<1/2$, as 
shown in the phase
diagram, Fig. 2. It is interesting to note that $\Delta=0$ on some 
line within the
coexistence phase, i.e. $\Delta$ changes sign continuously across 
this line. When
$u_0\approx v_0$ this line is $s=t$, as shown by the dashed line in 
Fig. 2. This line
is not a phase transition as far as the correlation $c(\bf{r})$ 
(Eq.~(\ref{46}) below) or
the critical currents are concerned. We expect, however, that the 
slow relaxation
phenomena, associated with the glass order, will disappear on this 
line.

 The boundary $s=1/2$ of the coexistence phase is a continuous 
transition with
$z\rightarrow 0$ at the boundary. On the other hand, the boundary at 
$t=1/2$ is a
discontinuous transition, $z+\Delta\rightarrow 0$ from the left while 
$\Delta=0,\,z\neq
0$ on the right, i.e. both $\Delta$ and $z$ are discontinuous.

To identify the various phases we consider the correlation function
\begin{equation}
c(r)=\langle\cos\varphi_{\alpha}({\bf 
r})\cos\varphi_{\alpha}(0)\rangle=
[\exp(-\phi_+)+\exp(-\phi_-)]/2
                                    \label{46}\end{equation}
where
\begin{equation}
\phi_{\pm}=\int_{1/L}^{\sqrt{\Delta_c}}q\,dq[1\pm 
J_0(qr)]G_{\alpha,\alpha}(q)/2\pi
                                       \nonumber\\
\end{equation}
and the system size $L$ appears as a low momentum cutoff. Using 
$G_{\alpha,\alpha}
(q)=\alpha(q)+\beta(q)+\gamma(q)$, the various correlations are 
summarized in Table
I. The ordered phases have finite correlation lengths defined as 
$\lambda_J=z^{-1/2}$
for the Josephson length, $\lambda_G=\Delta^{-1/2}$
for the glass correlation length and $\lambda'_G=(z+\Delta)^{-1/2}$ 
in the coexistence
phase. It is curious to note that in the coexistence phase 
$G_{\alpha,\alpha}$ has a
 $(2t-1)/(q^2+z+\Delta)$ term.
Since $z+\Delta\rightarrow 0$ much faster than
$2t-1\rightarrow 0$ at the boundary $t=1/2$, this leads to an 
apparent divergence of
$\lambda_G'$; however, $\phi_{\pm}$ is finite at $t\rightarrow 1/2$ 
and the transition is of first order.

\begin{table}
\[ \begin{array}{c|c|c|c}
\mbox{phase} & G_{\alpha,\alpha}(q) & 
c(L)\;;\;\;L<\lambda_J,\lambda_G &
c(L)\;;\;\;L>\min (\lambda_J,\lambda_G)  \\ \hline \hline
\mbox{Disorder} & \frac{8\pi(t+s)}{q^2} & \left(\frac{L}
{\lambda}\right)^{-4(t+s)} 
&\\ \hline \mbox{Josephson} 
& \frac{8\pi(t+s)}{q^2+z}-\frac{8\pi s z}{(q^2+z)^2} &
\left(\frac{L}{\lambda}\right)^{-4(t+s)} &
\left(\frac{\lambda_J}{\lambda}\right)^{-4(t+s)} \\ \hline
\mbox{Glass} & \frac{4\pi(1+2s)}{q^2}+\frac{4\pi(2t-1)}{q^2+\Delta} &
\left(\frac{L}{\lambda}\right)^{-4(t+s)} &
\left(\frac{L}{\lambda}\right)^{-2(1+2s)}\left(\frac{\lambda_G}
{\lambda}\right)^{-2(2t-1)} \\ \hline \mbox{Coexistence} &
\begin{array}[t]{l}\frac{4\pi(2t-1)}{q^2+z+\Delta}+\frac{4\pi(1+2s)}
{q^2+z} \\ -\frac{4\pi z}{(q^2+z)^2} \end{array} &
\left(\frac{L}{\lambda}\right)^{-4(t+s)} 
 & \left(\frac{\min (L,\lambda_G')}{\lambda}\right)^{-2(2t-1)}
\left(\frac{\min (L,\lambda_J)}{\lambda}\right)^{-2(1+2s)} \\ \hline
\end{array}    \]
\caption{ Correlations in junctions of size L; $c(L)$ determines $I_c$ 
via Eqs.~(\ref{48}, \ref{49}). }
\end{table} 
The phases with $z=0$ have power law correlations; for $L\rightarrow 
\infty$,
 $c(r)\sim r^{-4t-4s}$ in the disordered phase while $c(r)\sim 
r^{-2-4s}$ in the
glass phase. The glass phase leads to stronger decrease of $c(r)$ 
then what would
have been $c(r)$ in a disordered phase at $t<1/2$; a prefactor 
$(\lambda_J/\lambda)^{2(1-2t)}$ somewhat compensates for this 
reduction.

The phases with $z\neq 0$ have long range order. Note in particular 
the $z/(q^2+z)^2$
terms in $G_{\alpha,\alpha}$; these terms do not arise in RG since 
they are of higher
order in $z$ and are of interest away from the transition line. 
Note that in the Josephson phase $v_0\approx u_0$ is assumed,
so that $\sigma_0v_0\ll z$; otherwise the coefficient of 
$(q^{2}+z)^{-2}$ is modified.

 The correlation $c(L)$ measures the fluctuation effect on $\langle 
\cos\varphi_J\rangle$
in a finite junction, i.e. $\langle 
\cos\varphi_J\rangle\approx\sqrt{c(L)}$, which is
therefore related to the Josephson critical current $I_c$. The results 
for $c(L)$ are summarized in table I. Consider 
first a junction with
$L<\lambda_J$ (which is always the case in the $z$=0 phases). The 
current flows through
the whole junction and the system is equivalent to a point junction 
with an effective
Gibbs free energy, \begin{equation}
G_J^{eff}=E_J(L/\lambda)^2\sqrt{c(L)}\,\cos\varphi_J-(\phi_0/2\pi 
c)I^{ex}\varphi_J  .
                                           \label{47}\end{equation}
Here we assume (as at the end of section II) that point junction 
fluctuations can be ignored, i.e. $\phi_0 I_c /2c>T$ and the critical 
current of Eq.~(\ref{47}) can be deduced by its mean field equation 
(see section V for actual data). Thus,
the mean field value $I_{c1}^0$ (Eq.~(\ref{7a})) is reduced by the 
fluctuation factor, leading to a critical current
\begin{equation}
I_c=I_{c1}^0\sqrt{c(L)}    \mbox{\hspace{30mm}}   L<\lambda_J .
                                      \label{48}\end{equation}
For $L<\lambda_{J}, \lambda_{G}$ the parameters $\Delta$ and $z$ are no 
longer related to $\lambda_{J}$ or to $\lambda_{G}$; instead they are 
L dependent (Eq.~(\ref{34}) should be reevaluated leading to 
power laws of $L$). In particular $z$ affects $c(L)$ via the 
$(q^{2}+z)^{-2}$ terms by either a factor $\exp[2sz(L)L^{2}]$ (in the 
Josephson phase) or $\exp[z(L)L^{2}]$ (in the coexistence phase).
Although of unusual form, these factors are neglected in table I 
since $zL^{2}<1$. The dominant dependence in a small 
area junction, $L<\lambda_{J}, \lambda_{G}$ (for all phases) is a 
power law decrease of $c(L)$, leading to $I_{c}\sim L^{2-2t-2s}$.

 For systems with $L>\lambda_J$, the current flows in an area 
$~L\lambda_J$ near the
edges of the junction. The mean field value $I_{c2}^0$ 
(Eq.~(\ref{7a})) is reduced
now by a factor $\lambda_J^0/\lambda_J$. Using
$\langle\cos\varphi_J\rangle=\sqrt{c(L)}$ and
$z=u_0\langle\cos\varphi_J\rangle=1/\lambda_J^2$ we obtain 
$\lambda_J=\lambda_J^0
c^{-1/4}(L)$ with $\lambda_J^0=\lambda (\tau/8\pi E_J)^{1/2}$, as in 
section II. The
critical current is then
\begin{equation}
I_c=I_{c2}^0\sqrt[4]{c(L)}     \mbox{\hspace{30mm}}       L>\lambda_J
                                   \label{49}\end{equation}

The relevant range of temperatures $T\ll \tau$ (see section II), for 
typical junction parameters, is most of the range $T<T_{c}$, excluding 
only $T$ very close to $T_{c}$. Thus $t\ll 1$ and our main interest is 
the coexistence to glass transition at $s=\case {1}{2}$. This transition can be 
induced by a temperature change since $s=s(T)$ (see section III). Thus 
we consider $t\ll s$ for which $z\ll \Delta$ and $\lambda_{J}\gg 
\lambda_{G} \approx \lambda_{G}'$. When the transition at $s=\case {1}{2}$ is 
approached $\lambda_{J}$ diverges and for a given $L$ the system 
crosses into the regime $\lambda_{G}<L<\lambda_{J}$ (which includes 
the glass phase) where $c(L)\sim 
(L/\lambda)^{-4s}(\lambda_{G}/L)^{2}$ and $I_{c} \sim (L/\lambda)^
{1-2s}$. Since $L\gg \lambda$ we predict a sharp decrease of $I_{c}$ 
at some temperature $T_{J}$ for which $s(T_{J})=\case {1}{2}$; this is the 
finite size equivalent of the $L\rightarrow \infty$ phase transition.

\section{Discussion}

We have derived the effective free energy for a 2D josephson junction 
(Appendix A) and
studied it in presence of random magnetic fields. We show that a 
coupling between
replicas of the form $\cos(\varphi_{\alpha}-\varphi_{\beta})$ is 
essential for describing
the system. This coupling is generated by RG from the Josephson term 
in presence of the
random fields, or also from disorder in the Josephson coupling, a 
disorder whose finite
mean is $E_J$.

We find the phase diagram, Fig. 2, with four distinct phases defined 
in terms of a
Josephson ordering $z\sim\langle \cos
\varphi_J\rangle$ and a glass order parameter $\Delta$. At high 
temperatures thermal
fluctuations dominate and the system is disordered, $z=\Delta=0$. 
Lowering temperature at
weak disorder ($s<\case {1}{2}$) allows formation of a Josephson phase,
 $z\neq 0,\,\Delta=0$.
Further decrease of temperature leads by a first order transition to 
a coexistence phase
where both $z,\Delta\neq 0$. The Josephson and coexistence phases 
have similar diagonal
correlations (see table I). The main distinction between these phases 
is then the slow
relaxation times typical of glasses. Finally, at strong disorder and 
low temperatures the
glass phase with $z=0,\;\Delta\neq 0$ corresponds to destruction of 
the Josephson long
range order by the quenched disorder.

Our main result, relevant to experimental data with $t\ll 1$, is the 
coexistence to glass transition at $s=\case {1}{2}$. The critical behavior of 
$I_{c}(s)$ near this transition depends on the ratio $L/\lambda_{J}$; 
not too close to $s=\case {1}{2}$ where $L>\lambda_{J}$ we have from 
Eq.~(\ref{45}, \ref{49}) $\ln I_{c} \sim 1/(1-2s)$ while closer to 
$s=\case {1}{2}$ the divergence of $\lambda _{J}$ implies
 $L<\lambda _{J}$ with 
$I_{c} \sim (L/\lambda)^{1-2s}$. The junction ordering temperature 
$T_{J}$ corresponds to $s(T_{J})=\case {1}{2}$ so that either $\ln I_{c}\sim 
-(T_{J}-T)^{-1}$ (not too close to $T_{J}$) or $\ln I_{c} \sim 
(T_{J}-T)\ln L/\lambda$ close to $T_{J}$.

We reconsider now the experimental data 
\cite{Rogers,Virshup,Hashimoto,Sato,Strbik} where the junctions order
at temperatures well below the $T_c$ of the bulk. In our scheme, this 
can correspond to a
transition between the glass phase and the coexistence phase, a 
transition which may
occur even at low temperatures $t\ll 1$ provided $s$ decreases with 
temperature. As
discussed in section III, $s$ depends on a power of $\lambda$, in 
particular
$s\sim\lambda^2$ for short junctions, the experimentally relevant 
case. Thus $s$
decreases with temperature since $\lambda$ is temperature dependent. 
We propose then that
junctions with random magnetic fields (arising, e.g. from quenched 
flux loops in the
bulk) may order at temperatures well below $T_c$ of the bulk.

From critical currents \cite{Rogers,Virshup} at $4.2 K$ $I_c 
\approx 150-400 \mu A$ we infer $E_J \approx 1-4 K$ and $\lambda_J^0 
\approx 2-4 \mu m$, the latter is somewhat below the junction 
sizes $L\approx 5-50 
\mu m$. For the more recent data on YBCO junctions 
\cite{Hashimoto,Sato,Strbik} with $I_c \approx 0.4-6 mA$ we obtain 
$\lambda_J^0 \ll L$ and Eq.~(\ref{49}) applies. In fact, magnetic 
field dependence \cite{Sato} and $I_{c}\sim L$ dependence 
\cite{Umezawa} show directly that $\lambda_{J}<L$ is feasible.

We note also that mean field treatment of the effective free energy 
Eq.~(\ref{47}) is valid since thermal fluctuations of the effective 
point junction are weak (as assumed in sections II and IV), i.e. 
$\phi_0 I_c /2c >T$. E.g., at $80 K$ $\phi_0 I_c /2c =T$ corresponds 
to $I_c\approx 1\mu A$ while the mean field $I_c$ at the temperatures 
where $I_c$ disappears, i.e. at $0.4-0.8T_c$, should be comparable to 
its low temperature values 
\cite{Rogers,Virshup,Hashimoto,Sato,Strbik} of $I_c=0.1-6 mA$. Thus 
$\phi_0 I_c /2c \gg T$ and point junction type fluctuations can be 
neglected. 

Other interpretations of the data assume that the composition of the 
barrier material is affected by the superconducting material and 
becomes a metal \cite{Hashimoto} N or even a superconductor \cite{Strbik}
 S'. In an SNS junction the coherence length in the metal 
is temperature dependent and affects $I_c$ while the onset of an SS'S 
junction obviously affects $I_c$. Note, however, that the SNS 
interpretation with $\ln I_{c}\sim -T^{1/2}$ is consistent with the 
$T$ dependence but leads to an inconsistent value of the 
coherence length \cite{Hashimoto}. In our scheme, $\ln I_{c}\sim 
(T_{J}-T)\ln L/\lambda$ is consistent with the data \cite{Hashimoto} 
of the $100\times 100 \mu m^{2}$ junction showing a 
cusp in $I_{c}(T)$ near $T_{J}\approx 25 K$.
Further experimental data, and in 
particular the $L$ dependence of $I_c$, can determine the appropriate 
interpretation of the data.

The increasing research on large area junctions is motivated by 
device applications. The design of these junctions should consider 
the various types of disorder as studied in the present work. 
Furthermore,
 we believe that disordered large area junctions
deserve to be studied since they exhibit novel glass phenomena. In 
particular the
coexistence phase with both long range order and glass order is an
unusual type of glass.

\vspace{30mm}
{\bf Acknowledgments}: We thank S. E. Korshunov for valuable and 
stimulating discussions.
This research was supported by a grant from the Israel Science 
Foundation.

\newpage
\renewcommand{\theequation}{\thesection\arabic{equation}}
\appendix
\renewcommand{\theequation}{A\arabic{equation}}
\setcounter{equation}{0}
\section{Free energy of a 2D Josephson junction}

In this appendix we derive the effective free energy of a large area 
Josephson
junction. In Appendix A.1 boundary conditions and the Josephson phase 
are defined. In
Appendix A.2 the Gibbs free energy in presence of an external  
current is derived. In
Appendices A.3, A.4 the Gibbs free energy is derived explicitly for 
superconductors in
the Meissner state, i.e. no flux lines in the bulk; Appendix A.3 
considers long
junctions, i.e. $W>>\lambda$ (see Fig. 1) while Appendix A.4 
considers short ones,
$W<<\lambda$. Finally, in Appendix A.5 the free energy in presence of 
(quenched) flux
loops in the bulk is derived.

\subsection{Boundary conditions}

The barrier between the superconductors (region I in Fig. 1) is 
defined by allowing
currents $j_z(x,y)$ in the $z$ direction so that Maxwell's relation 
for the vector
potential ${\bf A}(x,y,z)$ is
\begin{equation}
{\bf \nabla}\times{\bf \nabla}\times{\bf A}=(4\pi/c)j_z\hat{z}
\label{A1}\end{equation}
where $\hat{z}$ is a unit vector in the $z$ direction.
There is no additional relation between $j_z$ and ${\bf A}$ (e.g. as 
in
superconductors). This allows $j_z$ to be a fluctuating variable in 
thermodynamic
averages.

Eq.~(\ref{A1}) implies that the magnetic field in the barrier ${\bf 
H}(x,y)={\bf
\nabla}\times{\bf A}$ is $z$ independent and $H_z=0$; thus the 
currents $j_x$,
$j_y=0$ as required.

Considering the superconductors in Fig. 1 we denote all 2D fields 
(i.e. $x$, $y$
components) at the right and left junction surfaces (i.e. $z=\pm 
d/2)$ with indices 1, 2,
respectively. Boundary conditions are derived \cite{Kulik} by 
integrating ${\bf
\nabla}\times{\bf A}$ around the dashed rectangle in Fig. 1, which 
since $j_y=0$, yields
continuity of the parallel magnetic fields
\begin{equation}
{\bf H}_1(x,y)={\bf H}_2(x,y)\,.
\label{A2}\end{equation}
Integrating ${\bf A}$ along the same rectangle yields for the vector 
potentials on
the junction surfaces,
\begin{equation}
A_{1x}-A_{2x}+\int_{-d/2}^{d/2}(\partial A_z/\partial x)\,dz=dH_y
\label{A3}\end{equation}
and a similar relation interchanging $x$ and $y$. Introducing the 
phases
$\varphi_i({\bf r})$, $i=1,2$ for the two superconductors and a gauge 
invariant
vector potential
\begin{equation}
{\bf A}_i'({\bf r})={\bf A}_i({\bf r})-(\phi_0/2\pi){\bf 
\nabla}\varphi_i({\bf r})
\label{A4}\end{equation}
yields for ${\bf A}_i'(x,y)$ on the junction surfaces
\begin{equation}
{\bf A}_1'(x,y)-{\bf A}_2'(x,y)=d{\bf 
H}(x,y)\times\hat{z}-(\phi_0/2\pi)
{\bf \nabla}\varphi_J(x,y)
\label{A5}\end{equation}
where $\varphi_J(x,y)$ is the Josephson phase,
\begin{equation}
\varphi_J(x,y)\equiv\varphi_1(x,y)-\varphi_2(x,y)-(2\pi/\phi_0)
\int_{-d/2}^{d/2}A_z\,dz
\label{A6}\end{equation}

\subsection{Gibbs free energy}

In the presence of a given external current ${\bf j}^{ex}$ passing 
through the
junction we separate the system into the sample with relevant 
fluctuations (e.g.
superconductors with barrier) and an external environment in which 
${\bf j}^{ex}$ is given.
Thermodynamic quantities are then given by a Gibbs free energy ${\cal 
G}({\bf H})$
where ${\bf H}$ is the field outside the sample which determines 
${\bf j}^{ex}$. The
situation which is usually studied is such that ${\bf j}^{ex}$ does 
not flow through
the sample \cite{deGennes} so that it is  uniquely defined 
everywhere. We need to
generalize this situation to the case in which ${\bf j}^{ex}$ flows 
through the
sample, a generalization which to our knowledge has not been developed
 previously.
   
In standard electrodynamics \cite{Landau}, in addition to the space 
and time dependent 
electric and magnetic fields ${\bf E}$ and ${\bf H}$, respectively, 
one defines a free
current ${\bf j}_f$, a displacement field ${\bf D}$ and an induction 
field ${\bf B}$ such
that
\begin{eqnarray}
{\bf \nabla}\times{\bf H}&=&(4\pi/c){\bf j}_f+(1/c)\partial{\bf 
D}/\partial t
\nonumber\\
{\bf \nabla}\times{\bf E}&=&-(1/c)\partial{\bf B}/\partial t
\label{A7}\end{eqnarray}
and only outside the sample ${\bf D}={\bf E}$, ${\bf B}={\bf H}$ 
and ${\bf j}_f={\bf j}^{ex}$. When the various electrodynamic fields 
change by a
small amount, the change in the sample's energy is the Poynting 
vector integrated
over the sample surface S (with normal $d{\bf s}$) in time $dt$
\begin{equation}
-dt\frac{c}{4\pi}\int_S{\bf E}\times{\bf H}\,d{\bf
s}=\int_V[\frac{1}{4\pi}{\bf H}\cdot d{\bf B}+\frac{1}{4\pi}{\bf 
E}\cdot d{\bf D}
+{\bf E}\cdot{\bf j}_f\,dt]\,dV
\label{A8}\end{equation}
where integration is changed from the surface $S$ to the enclosed 
volume $V$ by
Eq.~(\ref{A7}).
When ${\bf j}^{ex}$ does not flow through the sample, ${\bf j}_f=0$ 
and neglect of
${\bf D}$ (for low frequency phenomena) leads to the usual energy 
change \cite{deGennes}
$dE=\int{\bf H}\cdot d{\bf B}/4\pi$.

The general situation is described by keeping the surface integral in 
Eq.~(\ref{A8})
and in terms of the vector potential ${\bf A}$, where ${\bf 
E}=-(1/c)\partial{\bf A}\partial t$,

\begin{equation}
dE=\int_S\,d{\bf A}\times{\bf H}\,d{\bf s}/4\pi
\label{A9}\end{equation}
Thus the surface values of ${\bf A}$ and ${\bf H}$ (parallel to the 
surface)
determine the energy exchange $dE$ and there is no need to specify an 
${\bf H}$ or a
${\bf j}_f$ inside the sample, where in fact they are not uniquely 
determined.

Since ${\bf H}$ (on the surface) is determined by ${\bf j}^{ex}$ (via 
Eq.~(\ref{A7})
outside the sample) we define a Gibbs free energy ${\cal G}({\bf H})$ 
by a Legendre
transform

\begin{equation}
{\cal G}({\bf H})={\cal F}-(1/4\pi)\int_S {\bf A}\times{\bf H}\,
d{\bf s} 
\label{A10}\end{equation}
${\bf A}$ is determined now by a minimum condition
$\delta {\cal G}/\delta{\bf A}=0$ which indeed reproduces 
Eq.~(\ref{A9}).

We apply now Eq.~(\ref{A10}) to the Josephson junction system. We 
assume a time
independent current ${\bf j}^{ex}$, i.e. 
${\bf \nabla}\times{\bf H}=(4\pi/c){\bf j}^{ex}$ outside the sample 
and that the same
current ${\bf j}^{ex}$ flows through both superconductor-normal (SN) 
surfaces (e.g.
the superconductors close into a loop or that the current source is 
symmetric).
 Consider now the surface $S_1$ of
superconductor 1, which includes the superconductor-normal (SN) 
surface and the
superconductor-vacuum (SV) surface. The boundary of $S_1$ is a loop 
$J$ which encircles the
junction surface, oriented with normal ${+\hat z}$.  In terms of the 
gauge invariant
vector  ${\bf A}'={\bf A}-(\phi_0/2\pi){\bf \nabla}\varphi_1$, 
assuming ${\bf j}^{ex}$ 
is time independent, $\partial{\bf E}/\partial t=0$ and using
\[ {\bf \nabla}\varphi_1\times{\bf H}={\bf \nabla}\times(\varphi_1{\bf
H})-(4\pi/c)\varphi_1\,{\bf j}^{ex}\]
we obtain
\begin{equation}
\int_{S_1}{\bf A}\times{\bf H}\,d{\bf 
s}=\frac{\phi_0}{2\pi}[\oint_J\varphi_1{\bf
H}\cdot\,d{\bf l}-(4\pi/c)\int\varphi_1{\bf j}^{ex}\cdot d{\bf s}]+
\int_{S_1} {\bf A}'\times{\bf H}\cdot d{\bf s}
\label{A11}\end{equation}

The ${\bf j}^{ex}\cdot d{\bf s}$ term for both superconductors 
involves the difference
$\varphi_1-\varphi_2$ of the phases on the two SN surfaces. This 
difference
\cite{Schon} is related to the chemical potential difference in the 
external circuit
so that the corresponding term is $\varphi_J$ independent.

Consider next the insulator-vacuum (IV) surface. Since $H_z=0$ in the 
insulator only
the $A_zH_y$ or $A_zH_x$ terms contribute with
\begin{equation}
\int_{IV}{\bf A}\times{\bf H}\,d{\bf s}=-\oint_J{\bf H}\cdot\,d{\bf
l}\int_{-d/2}^{d/2}A_z\,dz
\label{A12}\end{equation}
Combining Eq.~(\ref{A11}), the similar term for superconductor 2 and
Eq.~(\ref{A12}), (ignoring $\varphi_J$ independent terms) we obtain,
\begin{equation}
{\cal G}({\bf H})={\cal F}-\frac{1}{4\pi}\int_{SV+SN}{\bf 
A}'\times{\bf H}\cdot d{\bf s}-
\frac{\phi_0}{8\pi^2}\oint\varphi_J{\bf H}\cdot\,d{\bf l} .
\label{A12a}\end{equation}

\subsection{Long superconductors}

We derive here an explicit free energy, in terms of the Josephson 
phase, for the
case $W>>\lambda_1, \lambda_2$ (see Fig. 1), where $\lambda_i$ 
(i=1,2) are the
London penetration lengths of the two superconductors, respectively. 
The incoming
current ${\bf j}^{ex}(x,y)$ is parallel to the $\hat{z}$ axis.

Consider the free energy \cite{deGennes} of
superconductor 1 (suppressing the subscript 1 for now)
\begin{equation}
{\cal F}=\frac{1}{8\pi}\int_{z\geq 
d/2}d^3r[\frac{1}{\lambda^2}(\frac{\phi_0}{2\pi}{\bf
\nabla}\varphi-{\bf A})^2+({\bf \nabla}\times{\bf A})^2]
\label{A13}\end{equation}

The superconductor is assumed to have no flux lines, i.e. 
$\varphi({\bf r})$ is
nonsingular. The vector ${\bf A}''={\bf A}-(\phi_0/2\pi){\bf 
\nabla}\varphi$ has
then 3 independent components (no gauge condition on ${\bf A}''$) and 
${\bf \nabla}\times{\bf A}''={\bf \nabla}\times{\bf A}$. The 
partition function
involves integration on all vectors ${\bf A}''$ and on its boundary 
values ${\bf
A}'_s({\bf r}_s)$ on the boundary ${\bf r}_s$ of the superconductor,
\begin{equation}
Z=\int{\cal D}{\bf A}'_s({\bf r}_s)\int{\cal D}{\bf A}''({\bf 
r})\exp[-{\cal F}\{{\bf
A}''({\bf r}), {\bf A}'_s({\bf r}_s\}/T] \, .
\label{A14}\end{equation}

We shift now the integration field from ${\bf A}''$ to $\delta{\bf 
A}'$ where 
${\bf A}''={\bf A}'+\delta{\bf A}'$ and ${\bf A}'$ is the solution of 
$\delta F/\delta{\bf A}'=0$, i.e.
\begin{equation}
{\bf \nabla}\times{\bf \nabla}\times{\bf A}'=-{\bf A}'/\lambda^2
\label{A15}\end{equation}
with ${\bf A}'={\bf A}'_s$ at the boundaries; thus $\delta {\bf 
A}'({\bf r}_s)=0$.
Since F is Gaussian, $F({\bf A}'+\delta{\bf A}')=F({\bf 
A}')+F(\delta{\bf A}')$ and
the integration on $\delta{\bf A}'$ is a constant independent of 
${\bf A}'_s({\bf r}_s)$. Thus
\[ Z\sim \int{\cal D}{\bf A}'_s({\bf r}_s)\exp[-{\cal F}\{{\bf 
A}'({\bf r})\}/T] \] where
\begin{equation}
{\cal F}\{{\bf A}'\}=\frac{1}{8\pi}\int d^3r[\frac{1}{\lambda^2}{\bf 
A}'^2+
{\bf \nabla}\times{\bf A}')^2]
\label{A16}\end{equation}
Note that Eq.~(\ref{A15}) implies ${\bf \nabla}\cdot{\bf A}'=0$ and 
therefore 
${\bf \nabla}^2{\bf A}'={\bf A}'/\lambda^2$. Note also that the 
currents obey ${\bf
j}=-(c/4\pi \lambda^2){\bf A}'$.

We are interested in boundary fields at the barrier which are 2D 
vectors, e.g. 
\[{\bf A}'_1(x,y)\equiv(A_{1x}'(x,y), A_{1y}'(x,y))\;.\] 
The effect of these fields
decays on a scale $\lambda$ so that for $z>>\lambda$, ${\bf 
A}'\sim{\hat
z}j^{ex}(x,y)$ also obeys London's equation 
$\lambda^2\nabla^2j^{ex}=j^{ex}$.
Therefore $j^{ex}$ is confined to a layer of thickness $\lambda$ near 
the SV surface. The solution for $z\geq d/2$ has the form
\begin{eqnarray}
&[A_x'({\bf r}), A_y'({\bf r})]={\bf A}_1'(x,y)\exp[-(z-d/2)/\lambda] 
\nonumber\\
&A_z'({\bf r})= \lambda {\bf \nabla}{\bf
A}_1'(x,y)\exp[-(z-d/2)/\lambda]-(4\pi\lambda^2/c)j^{ex}(x,y)
\label{A17}\end{eqnarray}
This ansatz is a solution of London's equation (\ref{A15}) provided 
that ${\bf A}_1'(x,y)$ is slowly varying on the scale of $\lambda$. 
The corresponding magnetic fields are
\begin{eqnarray}
({\bf \nabla}\times{\bf 
A})_x'&=&(1/\lambda)A_y'-(4\pi\lambda^2/c)\partial_yj^{ex}
+O(\nabla^2{\bf A}_1')    \nonumber\\
({\bf \nabla}\times{\bf 
A})_y'&=&-(1/\lambda)A_x'-(4\pi\lambda^2/c)\partial_xj^{ex}
+O(\nabla^2{\bf A}_1')
\label{A18}\end{eqnarray}
Since eventually ${\bf A}_1'\sim{\bf \nabla}\varphi_J$ 
(Eq~(\ref{A20}) below) we
evaluate ${\cal F}$ by neglecting terms with derivatives of ${\bf 
A}_1'$. Some care is, however, needed in evaluating cross terms 
with $j^{ex}$, which is not slowly
varying. Thus, $\int A_z'^2({\bf r})$ from Eq.~(\ref{A17}) involves
\[ \int j^{ex}{\bf \nabla}\cdot{\bf A}_1'\,dxdy=-\int{\bf 
A}_1'\cdot{\bf
\nabla}j^{ex}\,dxdy \] 
which cannot be neglected. Note that the line integral on the SV 
surface vanishes since on
this surface the perpendicular component of ${\bf A}_1'$ is zero, 
i.e. no currents flowing
into vacuum.  The $O(\nabla^2{\bf A}_1')$ terms in
(Eq.~(\ref{A18}) can be neglected since their product with $j^{ex}$ 
cannot be partially integrated without SV line integrals.

The cross terms from squaring Eqs.~(\ref{A17},\ref{A18}) involve
\[ \int[j^{ex}{\bf \nabla}\cdot{\bf A}_1'+{\bf A}_1'\cdot{\bf 
\nabla}j^{ex}]\,dxdy
=\int {\bf \nabla}\cdot(j^{ex}{\bf A}_1')\,dxdy=0  \,. \]

For superconductor 2 with $z<-d/2$ the solution has the form of 
Eq~(\ref{A17}) with 
${\bf A}_2'(x,y)$ replacing ${\bf A}_1'(x,y)$, the $z$ dependence has
$\exp[(z+d/2)/\lambda_2]$ and $-{\bf \nabla}\cdot{\bf A}'_2$ replaces 
${\bf \nabla}\cdot{\bf A}'_1$ in the equation for $A_z'$. For both 
superconductors (i=1,2), after $z$ integration, we obtain
\begin{equation}
{\cal F}_i=\int dx\,dy\,{\bf A}_i'^2(x,y)/8\pi\lambda_i + O(\partial{\bf 
A}_i')^2 .
\label{A19}\end{equation}.

Next we use the boundary conditions Eqs.~(\ref{A2}, \ref{A5}) to 
relate ${\bf A}_i'$ to
$\varphi_J$. Equations~(\ref{A2}, \ref{A18}) yield
\begin{equation}
{\bf A}_1'/\lambda_1-(4\pi\lambda_1^2/c){\bf \nabla}j_1^{ex}=
-{\bf A}_2'/\lambda_2-(4\pi\lambda_2^2/c){\bf 
\nabla}j_2^{ex}+O(\partial{\bf A}_i')^2
\end{equation}
while Eq.~(\ref{A5}) yields
\begin{equation}
{\bf A}_1'-{\bf A}_2'=d[-{\bf A}_1'/\lambda_1+(4\pi\lambda_1^2/c){\bf 
\nabla}j_1^{ex}]
-(\phi_0/2\pi){\bf \nabla}\varphi_J
\label{A20a}\end{equation}
Since ${\bf \nabla}j^{ex}$ is not slowly varying, the
ansatz Eq.~(\ref{A17}) is consistent (i.e. ${\bf A}'_i$ are slowly 
varying) only if the
junction is symmetric, $j_1^{ex}(x,y)=j_2^{ex}(x,y)$, 
$\lambda\equiv\lambda_1=\lambda_2$
and that the limit $d/\lambda\rightarrow 0$ is taken. Thus,
\begin{equation}
{\bf A}_1'=-{\bf A}_2'=-(\phi_0/4\pi)^2{\bf \nabla}\varphi_J
\label{A20}\end{equation}
The magnetic energy in the barrier is neglected since it involves 
$d/\lambda$.
The total free energy, from Eqs.~(\ref{A19},\ref{A20}) is then
\begin{equation}
{\cal F}={\cal F}_1+{\cal 
F}_2=\frac{1}{4\pi\lambda}\left(\frac{\phi_0}{4\pi}\right)^2
\int dxdy({\bf \nabla}\varphi_J)^2                                              
\label{A21}\end{equation}
If $j^{ex}=0$, Eqs.~(\ref{A19},\ref{A20}) are valid also for 
nonsymmetric
junctions and ${\cal F}$ has the form ~(\ref{A21}) with $2\lambda$ 
replaced by
$\lambda_1+\lambda_2+d$.

We proceed to find the Gibbs terms in (\ref{A12a}). Since 
Eq.~(\ref{A18}) and the
constraint of no current flowing into the vacuum, ${\bf 
A}'\times{\hat z}\cdot d{\bf l}=0$
yield ${\bf H}_{SV}=-(4\pi\lambda^2/c){\bf \nabla}j^{ex}\times{\hat 
z}$ on the SV surface, the loop integral becomes
\begin{equation}
\oint_J\varphi_J{\bf H}\cdot d{\bf 
l}=(4\pi\lambda^2/c)\oint_J\varphi_J{\bf
\nabla}j^{ex}\cdot d{\bf l}\times{\hat z}
\label{A22}\end{equation}

For the SV surface integration we use again ${\bf H}_{SV}$ so that for
superconductor 1,
\begin{eqnarray*}
&\int_{SV_1}{\bf A}'\times{\bf H}\cdot d{\bf 
s}=-(4\pi\lambda^4/c)\int_{SV_1}
A_z '{\bf \nabla}j^{ex}\cdot d{\bf s} \nonumber\\
&=(4\pi\lambda^2/c)\int{\bf A}_1'\cdot{\bf \nabla}j^{ex}\,dxdy 
+O(\nabla^2{\bf
A}_1', \varphi_J \mbox{ independent terms})
\end{eqnarray*}
where ${\bf \nabla}j^{ex}\cdot d{\bf s}$ is replaced by 
$\nabla^2j^{ex}\,dxdydz$
as $j^{ex}$ has dominant $x, y$ dependence. Using Eq.~(\ref{A20}) 
and adding
terms for both superconductors leads to
\[ \int_{SV}{\bf A}'\times{\bf H}\cdot d{\bf s}=\frac{2\phi_0}{c}
\int\varphi_J\,j^{ex}\,dxdy - \frac{\phi_0}{2\pi}\oint_J\varphi_J{\bf 
H}\cdot d{\bf l}   \]
Finally we obtain,
\begin{equation}
{\cal G}=\int 
dxdy\left[\frac{1}{4\pi\lambda}\left(\frac{\phi_0}{4\pi}\right)^2({\bf
\nabla}\varphi_J)^2 - \frac{\phi_0}{2\pi c}\varphi_J\,j^{ex}\right]
\label{A23}\end{equation}
Adding the Josephson tunneling term $\sim\cos\varphi_J$ leads to
Eqs.~(\ref{1},\ref{3}).

\subsection{Short superconductors}

 Consider superconductors with length $W_1, W_2 <<\lambda_1, 
\lambda_2$ (see
Fig. 1). The $\exp(-z/\lambda_1)$ in Eq.~(\ref{A17}) can be expanded 
to terms linear in $z$. Since now both $\exp(\pm z/\lambda_1)$ are 
allowed at  $z>0$, there
are two slowly varying surface fields ${\bf A}_1$, ${\bf H}_1$,
\begin{eqnarray}
[A_x',A_y'] &=&{\bf A}_1'(x,y)+z{\bf H}_1(x,y)\times{\hat z} +O(z^2) 
\nonumber\\
A_z'&=&A_{1z}'-z{\bf \nabla}\cdot{\bf A}_1' +O(z^2)
\label{A24}\end{eqnarray}
and the magnetic field is
\begin{equation}
{\bf H}={\bf H}_1(x,y)-(z/\lambda_1^2){\bf A}_1'(x,y)\times{\hat
z}+O(z^2,\partial A_z')
\label{A25}\end{equation}

The $x, y$ components of ${\bf H}={\bf H}_1^{ex}$ at $z=W_1$ define 
the boundary
conditions,
\begin{eqnarray}
H_{1x}-(W_1/\lambda_1^2)A_{1y}'&=&H_{1x}^{ex} \nonumber\\
H_{1y}+(W_1/\lambda_1^2)A_{1x}'&=&H_{1y}^{ex}
\label{A26}\end{eqnarray}
and similarly ${\bf H}_2^{ex}$ at $z=-W_2$.
\begin{eqnarray}
H_{2x}+(W_2/\lambda_2^2)A_{2y}'&=&H_{2x}^{ex} \nonumber\\
H_{2y}-(W_2/\lambda_2^2)A_{2x}'&=&H_{2y}^{ex}
\label{A27}\end{eqnarray}
Equations~(\ref{A26},\ref{A27}) and the boundary conditions 
(\ref{A2},\ref{A5})
at the junction determine all the fields ${\bf A}'_i, {\bf H}_i$ in 
terms of ${\bf
H}_i^{ex}$ and $\varphi_J$, e.g.

\begin{eqnarray}
A_{1x}'&=&\frac{\lambda_1^2}{\lambda_1^2W_2+\lambda_2^2W_1+dW_1W_2}
[(\lambda_2^2+W_2d)H_{1y}^{ex}
-\lambda_2^2H_{2y}^{ex}-W_2(\phi_0/2\pi)\partial_x\varphi_J] 
\nonumber\\
A_{2x}'&=&\frac{-\lambda_2^2}{\lambda_1^2W_2+\lambda_2^2W_1+dW_1W_2}
[(\lambda_1^2+W_1d)H_{2y}^{ex}
-\lambda_1^2H_{1y}^{ex}-W_1(\phi_0/2\pi)\partial_x\varphi_J] 
\nonumber\\
\nonumber\\
H_{1y}&=&\frac{\lambda_2^2W_1H_{2y}^{ex}+
\lambda_1^2W_2H_{1y}^{ex}+W_1W_2(\phi_0/2\pi)\partial_x\varphi_J}{
\lambda_1^2W_2+\lambda_2^2W_1+dW_1W_2}
\label{A28}\end{eqnarray}
 The boundary fields ${\bf H}_i^{ex}$ need to be slowly varying (of 
order ${\bf
\nabla}\varphi_J)$ so that Eq.~(\ref{A28}) is slowly varying; thus 
$H_z$, $A_{iz}\sim\nabla^2\varphi_J$ can be neglected.

The free energy (\ref{A16}), to leading order in $W_i/\lambda_i$ is
\begin{equation}
{\cal F}_1=(W_1/8\pi\lambda_1^2)\int{\bf 
A}_1'^2(x,y)\,dxdy+O((W_1/\lambda_1)^3({\bf
\nabla}\varphi_J)^2, (W_1/\lambda_1)^2{\bf \nabla}\varphi_J\cdot{\bf 
H}_1^{ex})
\label{A29}\end{equation}
Ignoring $\varphi_J$ independent terms,
\begin{eqnarray}
{\cal F}_1+{\cal F}_2=&\frac{\phi_0}{16\pi^2}\int
dx\,dy\left\{\frac{\phi_0}{2\pi} 
\frac{W_1(W_2\lambda_1)^2+W_2(W_1\lambda_2)^2}{(\lambda_1^2W_2+
\lambda_2^2W_1+dW_1W_2)^2}({\bf \nabla}\varphi_J)^2 \right. 
\nonumber\\  \\ 
&   \left. -2d\frac{\lambda_1^2W_2H_{1y}^{ex}
+\lambda_2^2W_1H_{2y}^{ex}}{(\lambda_1^2W_2+\lambda_2^2W_1+dW_1W_2)^2}\partial_x
\varphi_J+(x\leftrightarrow y) \right\}                                                    
\label{A30}\end{eqnarray} 
The free energy in the barrier 
\begin{equation}
{\cal F}_I=(d/8\pi)\int dxdy{\bf H}_1^2(x,y)
\label{A31}\end{equation}
precisely cancels the terms linear in ${\bf \nabla}\varphi_J$ in 
(\ref{A30}) so that
\begin{equation}
{\cal F}=\frac{1}{8\pi}\left(\frac{\phi_0}{2\pi}\right)^2
\frac{W_1W_2}{\lambda_1^2W_2+\lambda_2^2W_1}\int dxdy({\bf 
\nabla}\varphi_J)^2 .
\label{A32}\end{equation}

Considering next the Gibbs term, the SV surface involves $A_z'$ or 
$H_z$ which
are neglected. The SN surface involves ${\bf A}'(z=W_1)={\bf
A}_1+O(W_1^2\partial\varphi_J)$, hence
\begin{eqnarray}
-\frac{1}{4\pi}\int_{SN}{\bf A}'\times{\bf H}\cdot d{\bf
s}&=&\frac{\phi_0}{8\pi^2}\int dxdy\frac{\lambda_1^2W_2H_{1y}^{ex}
+\lambda_2^2W_1H_{2y}^{ex}}{\lambda_1^2W_2+\lambda_2^2W_1}
\partial_x\varphi_J-(x\leftrightarrow y) +...  \nonumber\\ 
&=&-\frac{\phi_0}{2\pi c}\int 
dx\,dy\,j^{ex}\varphi_J+\frac{\phi_0}{8\pi^2}\oint_J
\varphi_J{\bf H}\cdot d{\bf l}+...
\label{A33}\end{eqnarray}
where higher order terms in $W_i/\lambda_i$ and $\varphi_J$ 
independent terms are
ignored, and the fact that ${\bf H}\cdot{\bf l}$ is $z$ independent 
on the SV surface is used (this arises from zero current into the 
vacuum and neglecting
$H_z$). The current $j^{ex}$ is defined here as an average of the 
currents on both sides,
$j_i^{ex}=({\bf \nabla}\times{\bf H}_i^{ex})_z$ (which locally may 
differ), i.e.
\begin{equation}
j^{ex}=\frac{\lambda_1^2W_2j_1^{ex}+\lambda_2^2W_1j_2^{ex}}
{\lambda_1^2W_2+\lambda_2^2W_1}
\label{A34}\end{equation}
The Gibbs free energy is finally, 
\begin{equation}
{\cal G}={\cal F}-(\phi_0/2\pi c)\int dxdyj^{ex}(x,y)\varphi_J(x,y)
\label{A35}\end{equation}
with ${\cal F}$  given by Eq.~(\ref{A32}).

\subsection{Junctions with bulk flux loops}

Consider a junction with flux loops in the bulk of the 
superconductors. These
loops induce magnetic fields which couple to $\varphi_J$. To derive 
this coupling
we decompose the superconducting phase into singular $\varphi_s$ and 
nonsingular
$\varphi_{ns}$ parts, i.e.
\[ {\bf \nabla}\times{\bf \nabla}(\varphi_s+\varphi_{ns})=
{\bf \nabla}\times{\bf \nabla}\varphi_s\neq 0 \,. \] 
Define a 3 component vector ${\bf A}''={\bf A}-(\phi_0/2\pi){\bf
\nabla}\varphi_{ns}$ so that the free energy Eq.~(\ref{A13}) is
\begin{equation}
F=\frac{1}{8\pi}\int d^3r[\frac{1}{\lambda^2}(\frac{\phi_0}{2\pi}{\bf
\nabla}\varphi_s-{\bf A}'')^2+({\bf \nabla}\times{\bf A}'')^2]
\label{A36}\end{equation}
We shift the integration field ${\bf A}''$ by ${\bf 
A}''\rightarrow{\bf
A}''+\delta{\bf A}''$ (as in section A.3) where $\delta{\bf A}''=0$ 
at the boundaries and ${\bf A}''$ satisfies $\delta F/\delta{\bf A}''
=0$,  i.e.
\begin{equation}
{\bf \nabla}\times{\bf \nabla}\times{\bf A}''=
[(\phi_0/2\pi){\bf \nabla}\varphi_s-{\bf A}'']/\lambda^2
\label{A37}\end{equation}
Since ${\cal F}$ is Gaussian in ${\bf A}''$, the integration on 
$\delta{\bf A}''$
decouples from that of $\varphi_s$ and the boundary values. Define 
now ${\bf A}''={\bf A}'+{\bf A}_s$ where ${\bf A}_s$ is a specific 
solution of
Eq.~(\ref{A37}) and ${\bf A}'$ is the general solution of the 
homogeneous part of
(\ref{A37}), ${\bf \nabla}\times{\bf \nabla}\times{\bf A}'=-{\bf 
A}'/\lambda^2$,
which depends on boundary conditions, i.e. on $\varphi_J$.

Substituting Eq.~(\ref{A37}) for ${\bf A}_s$ in Eq.~(\ref{A36}) yields

\begin{equation}
{\cal F}=\frac{1}{8\pi}\int d^3r[\frac{1}{\lambda^2}(\lambda^2
{\bf \nabla}\times{\bf \nabla}\times{\bf A}_s-{\bf A}')^2+
({\bf \nabla}\times{\bf A}'+{\bf \nabla}\times{\bf A}_s)^2]  .
\label{A38}\end{equation}
In the absence of flux loops ${\bf \nabla}\times{\bf A}_s=0$ and 
Eq.~(\ref{A38})
reduces to the previous ${\cal F}({\bf A}')$ as in Eq.~(\ref{A16}). 
The terms which
depend only on ${\bf A}_s$ represent energies of flux loops in the 
bulk and
affect the thermodynamics of the bulk superconductors. Here we are 
interested in
temperatures well below $T_c$ of the bulk so that fluctuations of 
these flux loops are very slow and are then sources of frozen 
magnetic fields. The
thermodynamic average is done only on the boundary fields which 
determine 
${\bf A}'$, and are coupled to ${\bf A}_s$ by the cross terms in 
Eq.~(\ref{A38}),
\begin{eqnarray}
{\cal F}_s&=&(1/8\pi)\int d^3r[-2{\bf A}'\cdot{\bf \nabla}\times{\bf 
\nabla}\times{\bf A}'
+2{\bf \nabla}\times{\bf A}'\cdot{\bf \nabla}\times{\bf 
A}_s]\nonumber\\
&=&(1/4\pi)\int_S({\bf A}'\times{\bf \nabla}\times{\bf A}_s)\cdot 
d{\bf s}
\label{A39}\end{eqnarray}

The surface values of ${\bf A}'$ are determined by $\varphi_J$. The 
SV surface involves
$z$ integration of  ${\bf \nabla}\times{\bf A}_s$ with either 
$\exp(\pm z/\lambda)$,
Eq.~(\ref{A17}), or a linear function, Eq.~(\ref{A24}). In either 
case the randomness in 
${\bf \nabla}\times{\bf A}_s$ causes this integral to vanish. The 
relevant surface in Eq.~(\ref{A39}) is therefore the junction surface.

\renewcommand{\theequation}{B\arabic{equation}}
\setcounter{equation}{0}
\section{Hierarchical Replica Symmetry Breaking}

In this appendix we examine the full replica symmetry 
breaking formalism (RSB) and show that it reduces to 
the one step symmetry breaking solution, as studied in section IV. 
The method of RSB is
based \cite {Parisi,Mezard} on a representation of hierarchical matrices 
$A_{a b}$ in replica
space in terms of their diagonal $\tilde{a}$ and a one parameter function
$a(u)$, i.e. $ A_{a b} \rightarrow [ \tilde{a},a(u)] $. In our case
$ A_{a b}$ is related to the inverse Green's function
$ G^{-1}_{a b} $ which was obtained by Gaussian
Variational Method (GVM). 

To derive this representation, consider the 
 hierarchical form of a matrix $\hat{A}$,
\begin{equation}
\hat{A}= \sum_{i=0}^{k}a_{i}(\hat{C}_{i}-\hat{C}_{i+1}) +\tilde{a}\hat{I}
\label{B1}\end {equation}
Here $ \hat{C}_{i} $ is $ n \times n $ matrix whose nonzero elements 
are blocks of size $ m_{i}\times m_{i} $ along the diagonal; each 
matrix element within the blocks is equal to one; the last matrix 
equals the unit matrix $\hat{C}_{k+1}=\hat{I}$. The matrices
 $ \hat{C}_{i} $ satisfy relations which are useful for 
finding the representation of the product of matrices $ \hat{A}\hat{B} $  . 
Since the hierarchy is for $m_{i}/m_{i+1}$ integers, we have 
\begin {eqnarray}
\hat{C}_{i}& = &\sum_{j=i}^{k} (\hat{C}_{j}-\hat{C}_{j+1} )+\hat{I} 
\nonumber \\
\hat{C}_{i}\hat{C}_{j}& = &\left \{ \begin{array}{ll}
m_{i} \hat{C}_{j}    \mbox{\hspace{30mm}  $j\leq i$}, \nonumber \\
               m_{j} \hat{C}_{i}    \mbox{\hspace{30mm}   $j> i$} 
\end{array}        \right.              
                         \label{B2}\end{eqnarray}

The matrix product with a matrix $\hat{B}$,
\begin{equation}
\hat{B}= \sum_{i=0}^{k}b_{i}(\hat{C}_{i}-\hat{C}_{i+1}) +\tilde{b}\hat{I}
\label{B2a}\end {equation}
is found to be
\begin{equation}
\hat{A}\hat{B}= 
\sum_{j=0}^{k}(\hat{C}_{j}-\hat{C}_{j+1})[\sum_{i=j+1}^{k}(a_{i}b_{j}+
a_{j}b_{i})\,dm_{i}
-a_{j}b_{j}m_{j+1}+\sum_{i=0}^{j}a_{i}b_{i}\,dm_{i}]
+\hat{I}[\sum_{i=0}^{k}a_{i}b_{i}\,dm_{i}+\tilde{a}\tilde{b}]\,.                                   
\label{B3}\end{equation}
where $ dm_{i}=m_{i}-m_{i+1} $.

In the limit $n\rightarrow 0$ $m_i$ becomes a continuum variable $u$ in 
the range $0<u<1$ and $a_{i}$ becomes a function $a(u)$; thus the matrix 
$\hat{A}$ is represented
by $[\tilde{a},a(u)]$. The product of two matrices, using 
Eq.(\ref{B3}), becomes
 $\hat{A}\hat{B}\rightarrow [\tilde  {c}, c(u)]$ where 
\begin{eqnarray} 
\tilde {c} &=&\tilde{a}\tilde{b}-<ab>      
\nonumber\\                            
 c(u)  &=& ( \tilde{a}-<a>)b(u)+( \tilde{b}-<b>)a(u)-\int_{0}^{u}
[a(u)-a(v)][b(u)-b(v)] d v 
                                  \label{B4}\end{eqnarray}
and $ <a>= \int_{0}^{1} a(u) d u$.

To find the inverse $\hat{B}=\hat{A}^{-1}$ we solve for $\tilde  
{c}=1$, $c(u)=0$ and find
\begin{equation}
\tilde{b} - b(u)  = \frac {1} {u ( \tilde {a}-<a>-[a](u) ) } - \int 
_{u}^{1} \frac {d v} { v^{2} (\tilde {a}-< a >-   [a](v))}
                                      \label{B5}\end{equation}

\begin{equation}
\tilde{b}  = \frac {1} {\tilde {a}-<a>}[1-\int_{0}^{1} \frac {d 
v\,[a](v)}{v^{2}(\tilde {a}-<a>-[a](v))}-\frac { a(0)} {\tilde 
{a}-<a>}] 
                                \label{B6}\end{equation}

\begin{equation}
[a](u) \equiv u a(u)-\int_{0}^{u}d v\,a(v) \, .
                                          \label{B7}\end{equation}

The inverse Green's function is from Eq.~(\ref{26})  
\begin {equation}
4\pi G_{ab}^{-1}(q)=\frac{1} {2t}[\delta_{ab} 
(z+q^{2})-v_{0}\sigma_{ab}-q^{2} s/t]
                                              \label{B8}\end{equation}
which for $\hat{\sigma}\rightarrow[\tilde{\sigma}, \sigma(u)]$ 
parameterizes as $[\tilde{a}_{q}, a_{q}(u)] $ with
\begin{eqnarray}
\tilde{a}_{q} & = & \frac{1}{2t} [q^{2} (1-s/t) + z 
- v_{0} \tilde {\sigma}] \nonumber \\   
a_{q}(u) & = & -\frac{1} {2t}[q^{2}s/t+v_{0}\sigma(u)]
                               \label{B9}\end{eqnarray}

Since the sum on each row of $ \hat{\sigma} $ vanishes 
(Eq.~(\ref{27}) we obtain $ \tilde {\sigma}= 
<\sigma>$. Therefore the denominator under the integration in 
Eqs.~(\ref{B5},\ref{B6}) assumes the form 
\begin {equation}
\tilde{a}_{q}-<a_{q}> - [a_{q}](u) = \frac {1} {2t} (q^{2} + z+\Delta 
(u))
                                           \label{B10}\end{equation}
where the order parameter $\Delta(u)$ is defined by
 $ \Delta (u)=v_0[\sigma](u) $. From Eq.~(\ref{B5}) the representation
of the Green's
function takes the form $ (4\pi)G_{ab}\rightarrow [\tilde 
{b}_{q}, b_{q}(u)]$ with
\begin {equation}
\tilde{b}_{q}-b_{q}(u) = 2t[\frac {1} {u[q^{2} + z+\Delta (u)]}
-\int_{u}^{1} \frac 
{dv}{v^{2}[q^{2} + z+\Delta (v)] }]\,.
                                     \label{B12}\end {equation}
The GVM equation for $ \sigma(u)$ is from  Eq.~(\ref{27}) 
$ \sigma(u) = \exp[-B(u)]$, where from Eq.~(\ref{24})
\begin {equation}
B(u)=4\pi \int \frac {d^{2}q} {(2\pi)^{2}}(\tilde{b}_{q}-b_{q}(u)) = 
\frac {g(u)} {u} - \int_{u}^{1} \frac {dv\, g(v)} {v^{2}}  \, .
                                           \label{B13}\end {equation}
Eq.~(\ref{B12}), after summation on $q$, identifies
\begin {equation}
g(u)=2t \log \frac{\Delta_{c}}{\Delta(u) +z } 
                                           \label{B14}\end {equation}
Using $\sigma '(u)=d [\exp (-B(u)/2)]/du=
-\sigma (u)g'(u)/u$ and the definition of $\Delta(u)$, 
$\Delta '(u)=v_{0}u\sigma '(u)$ we obtain                                           
\begin {equation}
\frac{\Delta '(u)}{u}=-\frac{d} 
{du}[\frac{\Delta '(u)}{g'(u)}] 
                                         \label{B17}\end {equation}
which from Eq.~(\ref{B14}) can be written as
\begin{equation}
\left(\frac{1}{u}-\frac{1}{2t}\right)\frac{d\Delta}{du}=0\,.
                                  \label{B18}\end {equation}
The solution of this equation is a step function, i.e. 
$\Delta(u)$ jumps from zero to
a constant value at $u=2t$, which is precisely the one step solution.

We note that keeping finite cutoff corrections \cite{Korshunov} 
spoils this
correspondence. The variational method is, however, designed for weak 
coupling systems
and an infinite cutoff procedure is approporiate.

\newpage

\begin{center}
\LARGE{Erratum: Disorder in two-dimensional Josephson junctions}
\footnote{Phys. Rev. B, Jan. 1998}
\newline
\Large{[Phys. Rev. B {\bf 55}, 14499 (1997)]}
\end{center}
\vspace{5mm}
\begin{center}
\large{Baruch Horovitz and Anatoly Golub}
\end{center}

\vspace{10mm}

One of the regions in our disorder-temperature ($s$-$t$) phase diagram 
had a negative glass order parameter $\Delta$, coexisting with a 
finite renormalized Josephson coupling $z$; this region was 
$s<t<\case{1}{2}$ (see Fig. 2). While this is a formal solution of the 
replica symmetry breaking equations, we have realized now that this 
solution is unstable.

The average probability distribution of the Josephson phase 
$|\varphi_J (q)|^2$ is given 
by $\sim \exp [-|\varphi_J (q)|^2 /2G_{\alpha ,\alpha}(q)]$ \newline 
[ M. M\'{e}zard
and G. Parisi, J. Phys. I (France) {\bf 1}, 809 (1991), Appendix III]
where $G_{\alpha ,\alpha}(q)$ is the replica diagonal Green's function. 
Thus a thermodynamic stability condition is that $G_{\alpha 
,\alpha}(q)>0$ for all $q$. In the coexistence phase we obtain 
(correcting a minor error in the entry for ``coexistence'' in table I)
\[ G_{\alpha ,\alpha}(q)= 
\frac{4\pi (2t-1)}{q^2+z+\Delta}+\frac{4\pi (1+2s)}{q^2+z}+
\frac{4\pi z(1-2s)}{(q^2+z)^2} \, . \]
For $\Delta >0$ we have $G_{\alpha ,\alpha}(q)>0$ for all $q$ and
the coexistence phase is valid for $t<s<\case{1}{2}$. 
However, for $\Delta <0$ the minimum of $G_{\alpha ,\alpha}(q)$ is 
 at $q=0$ 
and $G_{\alpha ,\alpha}(0)>0$ yields the stability condition 
$1-2t<2(z+\Delta)/z$. From Eq. (46) we have
\[\frac{z+\Delta}{z}=e\left (\frac{2tv_0}{u_0}\right )^{2/(1-2s)}
\left (\frac{2tv_0}{\Delta_c}\right )^{2(t-s)/[(1-2t)(1-2s)]}\, ,\]
i.e. for weak coupling $v_0\ll \Delta_c$ and with $v_0/u_0 \sim O(1)$  
this shows $z+\Delta \ll z$ for all $s<t<\case{1}{2}$, unless $t-s$ is 
very small, of order $1/ln(\Delta_c/2tv_0)$. Thus, at $t=s$, up to 
nonuniversal $1/ln(\Delta_c/2tv_0)$ terms, the coexistence phase 
becomes unstable and for $s<t<\case{1}{2}$ is replaced by the 
Josephson phase where $\Delta=0$,
$z\neq 0$. The phase boundary between the coexistence phase and the 
Josephson phase is therefore a continuous phase transition at the 
dashed line in Fig. 2, i.e. $s=t$, $s<\case{1}{2}$ (rather than a 
first order transition at the vertical line $t=\case{1}{2}$, 
$s<\case{1}{2}$). All other conclusions in the paper remain intact.

\end{document}